\documentclass[preprint,superscriptaddress,prb]{revtex4}

\usepackage{graphicx}
\usepackage{epstopdf}
\usepackage{bbm}
\usepackage{amsmath}
\usepackage{eqnarray}
\usepackage{xcolor}

\begin{document}

\title{Mesoscopic Klein-Schwinger effect in graphene}

\author{A. Schmitt}\email{aurelien.schmitt@phys.ens.fr}
\affiliation{Laboratoire de Physique de l'Ecole normale sup\'erieure, ENS, Universit\'e
PSL, CNRS, Sorbonne Universit\'e, Universit\'e de Paris, 24 rue Lhomond, 75005 Paris, France}
\author{P. Vallet}
\affiliation{Laboratoire Ondes et Mati\`ere d'Aquitaine, 351 cours de la lib\'eration, 33405 Talence, France}
\author{D. Mele}
\affiliation{Laboratoire de Physique de l'Ecole normale sup\'erieure, ENS, Universit\'e
PSL, CNRS, Sorbonne Universit\'e, Universit\'e de Paris, 24 rue Lhomond, 75005 Paris, France}
\affiliation{Univ. Lille, CNRS, Centrale Lille, Univ. Polytechnique Hauts-de-France, Junia-ISEN, UMR 8520-IEMN, F-59000 Lille, France.}
\author{M. Rosticher}
\affiliation{Laboratoire de Physique de l'Ecole normale sup\'erieure, ENS, Universit\'e
PSL, CNRS, Sorbonne Universit\'e, Universit\'e de Paris, 24 rue Lhomond, 75005 Paris, France}
\author{T. Taniguchi}
\affiliation{Advanced Materials Laboratory, National Institute for Materials Science, Tsukuba,
Ibaraki 305-0047,  Japan}
\author{K. Watanabe}
\affiliation{Advanced Materials Laboratory, National Institute for Materials Science, Tsukuba,
Ibaraki 305-0047, Japan}
\author{E. Bocquillon}
\affiliation{Laboratoire de Physique de l'Ecole normale sup\'erieure, ENS, Universit\'e
PSL, CNRS, Sorbonne Universit\'e, Universit\'e de Paris, 24 rue Lhomond, 75005 Paris, France}
\affiliation{II. Physikalisches Institut, Universit\"at zu K\"oln, Z\"ulpicher Strasse 77, 50937 K\"oln}
\author{G. F\`eve}
\affiliation{Laboratoire de Physique de l'Ecole normale sup\'erieure, ENS, Universit\'e
PSL, CNRS, Sorbonne Universit\'e, Universit\'e de Paris, 24 rue Lhomond, 75005 Paris, France}
\author{J.M. Berroir}
\affiliation{Laboratoire de Physique de l'Ecole normale sup\'erieure, ENS, Universit\'e
PSL, CNRS, Sorbonne Universit\'e, Universit\'e de Paris, 24 rue Lhomond, 75005 Paris, France}
\author{C. Voisin}
\affiliation{Laboratoire de Physique de l'Ecole normale sup\'erieure, ENS, Universit\'e
PSL, CNRS, Sorbonne Universit\'e, Universit\'e de Paris, 24 rue Lhomond, 75005 Paris, France}
\author{J. Cayssol}
\affiliation{Laboratoire Ondes et Mati\`ere d'Aquitaine, 351 cours de la lib\'eration, 33405 Talence, France}
\author{M. O. Goerbig}
\affiliation{Laboratoire de Physique des Solides, CNRS UMR 8502, Univ. Paris-Sud, Universit\'e
Paris-Saclay, F-91405 Orsay Cedex, France}
\author{J. Troost}
\affiliation{Laboratoire de Physique de l'Ecole normale sup\'erieure, ENS, Universit\'e
PSL, CNRS, Sorbonne Universit\'e, Universit\'e de Paris, 24 rue Lhomond, 75005 Paris, France}
\author{E. Baudin} \email{emmanuel.baudin@phys.ens.fr}
\affiliation{Laboratoire de Physique de l'Ecole normale sup\'erieure, ENS, Universit\'e
PSL, CNRS, Sorbonne Universit\'e, Universit\'e de Paris, 24 rue Lhomond, 75005 Paris, France}
\author{B. Pla\c{c}ais} \email{bernard.placais@phys.ens.fr}
\affiliation{Laboratoire de Physique de l'Ecole normale sup\'erieure, ENS, Universit\'e
PSL, CNRS, Sorbonne Universit\'e, Universit\'e de Paris, 24 rue Lhomond, 75005 Paris, France}

\begin{abstract}
\textbf{Strong electric field annihilation by particle-antiparticle pair creation, also known as the Schwinger effect, is a non-perturbative prediction of quantum electrodynamics. Its experimental demonstration remains elusive, as threshold electric fields are extremely strong and beyond current reach. Here, we propose a mesoscopic variant of the Schwinger effect in graphene, which hosts Dirac fermions with an approximate electron-hole symmetry. Using transport measurements, we report on universal 1d-Schwinger conductance at the pinchoff of ballistic graphene transistors. Strong pinchoff electric fields are concentrated within approximately 1 $\mu$m of the transistor's drain, and induce Schwinger electron-hole pair creation at saturation. This effect precedes a collective instability toward an ohmic Zener regime, which is rejected at twice the pinchoff voltage in long devices. These observations advance our understanding of current saturation limits in ballistic graphene and provide a direction for further quantum electrodynamic experiments in the laboratory.}
\end{abstract}

\maketitle

A variety of important physical phenomena requires a non-perturbative understanding  of quantum field theory. This includes solitonic waves, but also several depply quantum mechanical phenomena like the confinement of quarks in quantum chromodynamics. While many non-perturbative problems are notably hard to compute, some are within reach and yield accurate predictions. Among the most striking non-perturbative predictions of quantum field theory is the instability of an electric field under the creation of particle-antiparticle pairs. The Schwinger effect (SE) states that pairs are created, out of a false vacuum with an electric field, to minimize energy. It is a simple yet non-trivial and non-perturbative prediction of quantum electrodynamics [\onlinecite{Sauter1931zphys,Heisengerg1936zphys,Schwinger1951prb,Itzykson2006mcgrawhill}]. The pair-creation rate per unit $d$-dimensional volume reads
$w(E)\propto \sum_{n\geq1}\left(\frac{E}{n}\right)^{\frac{d+1}{2}}e^{-\pi \frac{nE_S}{E}}$ (see prefactors in [\onlinecite{Itzykson2006mcgrawhill}] and Supplementary Section-I), where $E$ is the electric field and $E_S=\Delta_S^2/e\hbar c=1.32\; 10^{18}\;\mathrm{Vm^{-1}}$, called Schwinger field,  corresponds to the electron-mass energy $\Delta_S=mc^2=511\;\mathrm{keV}$, $2\Delta_S$ being the threshold energy needed to create an electron-positron pair at a characteristic length-scale given by the Compton length $\lambda_C=hc/\Delta_S$.
Large efforts have been devoted to produce such extremely strong electric fields in the laboratory to check this prediction explicitly [\onlinecite{Schutzhold2008prl}].
Unfortunately, the attempts to observe SE in the last decades have not yet been met with success.

With the advent of graphene, a novel playground for the study of relativistic effects has been opened in the completely different framework of condensed matter physics [\onlinecite{Gavrilov1996prd,Shytov2009ssc,Schutzhold2009asl,Dora2010prb,Gavrilov2012prd,Katsnelson2012jetp}]. In such a mesoscopic variant, electron-positron pairs are substituted by electron-hole pairs, speed of light $c$ by the Fermi velocity $v_F\simeq10^6\;\mathrm{m/s}$, and
the rest energy by a bandgap  (see Section \ref{Schwinger-voltage} below).
While graphene is intrinsically gapless, we show below  that an effective gap is provided for 1d-transport.
Our experiments point that this effective energy scale $\Delta_S$ is less than $0.2\;\mathrm{eV}$, in which case the Schwinger fields $E_S=\Delta_S^2/e\hbar v_F\lesssim6.10^{7}\;\mathrm{Vm^{-1}}$ are experimentally accessible, while remaining smaller than breakdown fields of the embedding hBN dielectric [\onlinecite{Pierret2022MatRes}].
Neutral single layer graphene, considered in Refs.[\onlinecite{Shytov2009ssc,Dora2010prb,Gavrilov2012prd}], corresponds to the gapless 2d limit
where $E_{S}=0$ so that the 2d-Schwinger rate  reduces to a super-linear  $J\propto E^{\frac{3}{2}}$ current-field relation. This power law is indeed observed in single layer graphene [\onlinecite{Vandecasteele2010prb}], but alternatively interpreted in terms of Zener tunneling. More recently, the same dependence has been reported in twisted bilayer graphene heterostructures, where the SE develops on top of a  saturation current
[\onlinecite{Berdyugin2022science}],  with a sign reversal of the Hall effect as an additional signature.  Our experiment is motivated by the study of transport in bottom gated (gate voltage $V_g$) single layer graphene field-effect transistors, where pair creation is subject to a finite breakdown field at large doping.
Most saliently, our experiment investigates the large bias regime where a giant Klein collimation establishes a quasi-one dimensional transport characterized by a transport gap, and where the Schwinger effect is accurately fitted by the 1d pair-creation rate with a finite Schwinger bandgap  $\Delta_S$. It develops over a ballistic junction, of length $\Lambda< 1\;\mathrm{\mu m}$ set by the gate dielectric thickness (see Section \ref{Schwinger-voltage}), which builds up at current saturation. This saturation is due to a large drain-source voltage $V$ that reduces the drain-gate voltage $V-V_g$, inducing a suppression of the electronic density $n_d$ at the drain below the channel density. The latter is given by the density at the source  $n_s$ (and the chemical potential at the source $\mu_s$), itself set by the gate voltage $V_g$.
This so-called pinchoff generates a peak effect with a large local electric field prone to the ignition of Schwinger-pair creation which shows up as the breakdown of the giant Klein collimation. As this situation differs from the canonical vacuum breakdown, we distinctively call it the Klein-Schwinger effect (KSE).

\begin{figure}[h!!!!]
\centerline{\includegraphics[width=16cm]{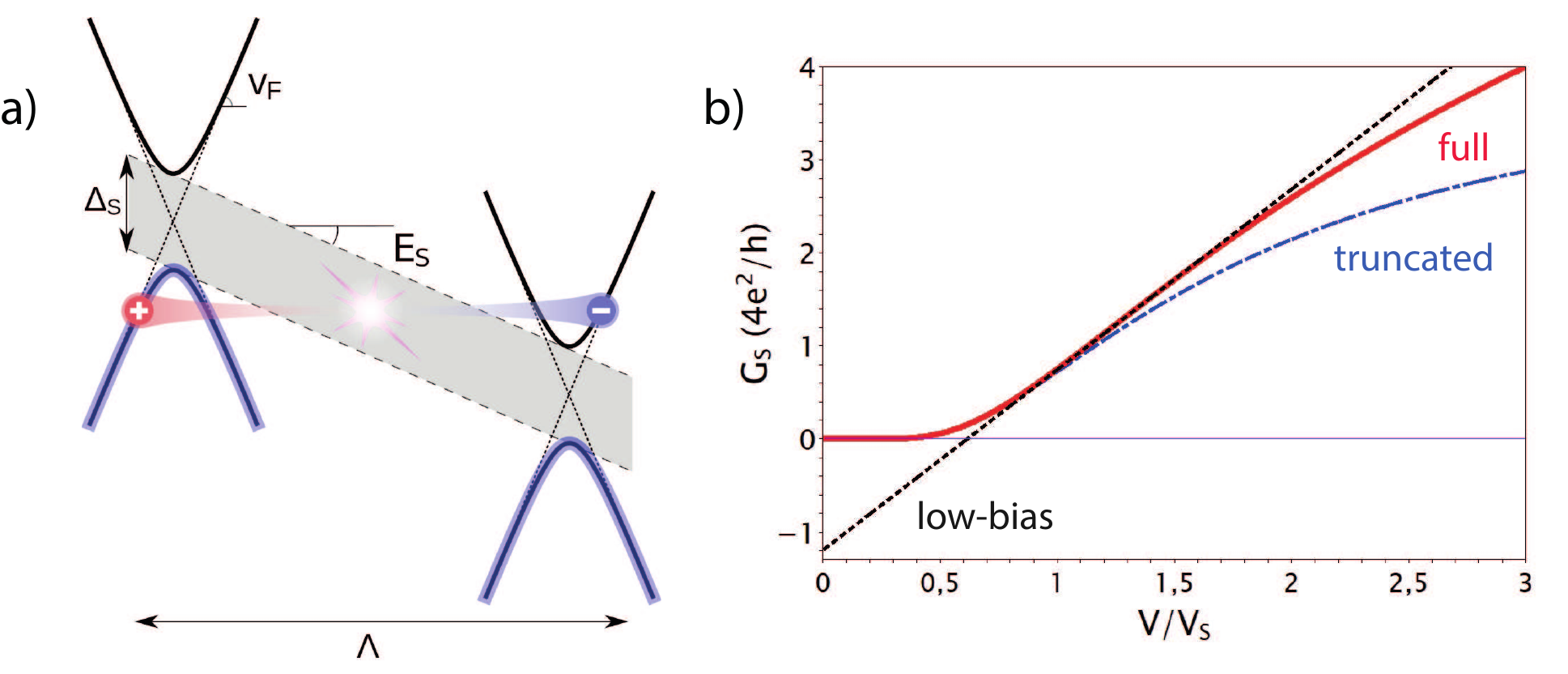}}
\caption{Mesoscopic Schwinger  effect for massive 1d-Dirac fermions.
Panel a: Sketch of the 1d-Schwinger effect across a graphene junction of length $\Lambda$ and gap $\Delta_S$, defining a Schwinger field  $E_S=\Delta_S^2/e\hbar v_F$ and a junction Schwinger voltage  $V_S=E_S\Lambda$. It leads to the proliferation of electron-hole pairs at a rate $w_{1d}\Lambda$, with $w_{1d}$ given by Eq.(\ref{1d-rate}) and a pair-current given by Eq.(\ref{1d-Schwinger-current}).
Panel b: the differential conductance $G_S=\partial I_S/\partial V$ given by  Eq.(\ref{1d-Schwinger-conductance}) for  $g_s=g_v=2$ (red line). It is well approximated by the affine approximation (\ref{1d-Schwinger-conductance-dl}), $G_S=1.20\left(\frac{V}{V_C}-1\right)\times\frac{4e^2}{h}$, with $V_C=0.62V_S$  (black dashed line). The truncated Schwinger conductance (blue dash-dotted line), $G^*_S=4g_sg_v\frac{e^2}{h}\left(1+\pi\frac{V_S}{V}\right)e^{-\pi \frac{V_S}{V}}$ (see main text), significantly deviates from the full series  (\ref{1d-Schwinger-conductance}).
 }
\label{Figure1_mesoscopic_Schwinger}
\end{figure}

Let us detail theoretical prediction for the mesoscopic 1d-Schwinger effect, as sketched in Fig.\ref{Figure1_mesoscopic_Schwinger}-a. Pair creation is monitored by the current generated across the junction upon dissociation in the large electric field.
Starting from the 1d-Schwinger rate in Eq.(\ref{1d-rate}), and taking into account the factor $2$ for pair dissociation as well as spin and  valley degeneracies $g_s$ and $g_v$, the pair current $I_{S}$ and differential conductance $G_{S}$ can be written as :
\begin{eqnarray}
w_{1d}&=&\frac{2e}{h}E\sum_{n=1}^\infty\frac{e^{-n\pi \frac{E_S}{E}}}{n}=\frac{2e}{h}E \ln{\left(\frac{1}{1-e^{-\pi \frac{E_S}{E}}}\right)} \label{1d-rate}\\
I_{S}&=&4V \ln{\left(\frac{1}{1-e^{-\pi \frac{V_S}{V}}}\right)}\times g_sg_v\frac{e^2}{h}\qquad\mathrm{with}\qquad V_{S}=\Lambda E_{S}=\Lambda\frac{\Delta_S^2}{e\hbar v_F}
\label{1d-Schwinger-current}\\
G_{S}&=&4\left[\frac{\pi\frac{V_S}{V}}{e^{\pi \frac{V_S}{V}}-1}+\ln{\left(\frac{1}{1-e^{-\pi \frac{V_S}{V}}}\right)}\right]\times g_sg_v\frac{e^2}{h}\label{1d-Schwinger-conductance}\\
&\approx&1.20 \left(\frac{V}{0.62 V_S}-1\right)\times g_sg_v\frac{e^2}{h}\qquad\qquad (V_C=0.62 V_S\lesssim V\lesssim 2V_{S})
\label{1d-Schwinger-conductance-dl} \end{eqnarray}
where $V$ is the total bias and $V_{S}$ the Schwinger voltage, product of the Schwinger field $E_{S}=\frac{\Delta_S^2}{e\hbar v_F}$ by the length  $\Lambda$ of the pair-creation domain. The differential conductance (\ref{1d-Schwinger-conductance}) possesses an inflection point at $V=1.22 V_S$, supporting an affine approximation given by Eq.(\ref{1d-Schwinger-conductance-dl}) and characterized by a critical voltage  $V_C=0.62 V_S$ and a zero-bias extrapolate $-G_0$ stemming from a quantized pair conductance  $G_0/g_sg_v=2\times0.60\;\frac{e^2}{h}$ (per spin and valley), with the prefactor $2$ as a tag for pair dissociation. The Schwinger conductance (\ref{1d-Schwinger-conductance}) is plotted in Fig.\ref{Figure1_mesoscopic_Schwinger}-b for 1d-graphene where $g_s=g_v=2$ (solid red line), including the affine approximation (dotted black line for Eq.(\ref{1d-Schwinger-conductance-dl})) with  $G_0=0.186\;\mathrm{mS}$. Also shown in the figure is the conductance $G^*_S=4g_sg_v\frac{e^2}{h}\left(1+\pi\frac{V_S}{V}\right)e^{-\pi \frac{V_S}{V}}$ (dash-dotted blue line) calculated with the Schwinger rate (\ref{1d-rate}) truncated to its first  $n=1$ term; the truncated variant deviates from the full Schwinger conductance for $V/V_S\gtrsim1.5$ and $G_S\gtrsim0.2\;\mathrm{mS}$. Comparison of massive 1d and 2d-Schwinger conductances, as well as that of 3d-Schwinger and the non relativistic Fowler-Nordheim mechanism, are contrasted as discussed in Supplementary Section-I.

Measurements are performed at room temperature in a series of $6$ hBN-encapsulated single-layer-graphene transistors deposited on local back-gates, made of graphite for devices GrS1-3 and gold for devices  AuS1-3, and equipped with high-transparency edge contacts (see Methods and Supplementary Section Table-I). As explained below (Section \ref{Ballistic pinchoff}), high-mobilities $\mu\gtrsim 10\;\mathrm{m^2V^{-1}s^{-1}}$, large dimensions $(L, W)\gtrsim10\;\mathrm{\mu m}$, high doping $n_s$ and small gate-dielectric thicknesses $t_{hBN}\lesssim 100\;\mathrm{nm}$, are key ingredients for observing the Schwinger effect. High mobility provides current saturation at low bias $V_{sat}<V_g$, whereas high-doping and long channels are needed to reject the Zener channel instability at a large bias $V_Z>V_g$. The Schwinger effect is visible in the bias window $[V_{sat},V_Z]$ where Klein-collimation is effective, at a Schwinger voltage $V_S\sim V_g$ controlled by the channel doping $n_s$ and dielectric thickness $t_{hBN}$.  Data reported below refer mostly to the representative GrS3 sample, leaving the full sample series description for the Supplementary Section-IV.

\section{Klein collimation junction}\label{Ballistic pinchoff}

\begin{figure}[h!!!!]
\centerline{\includegraphics[width=17cm]{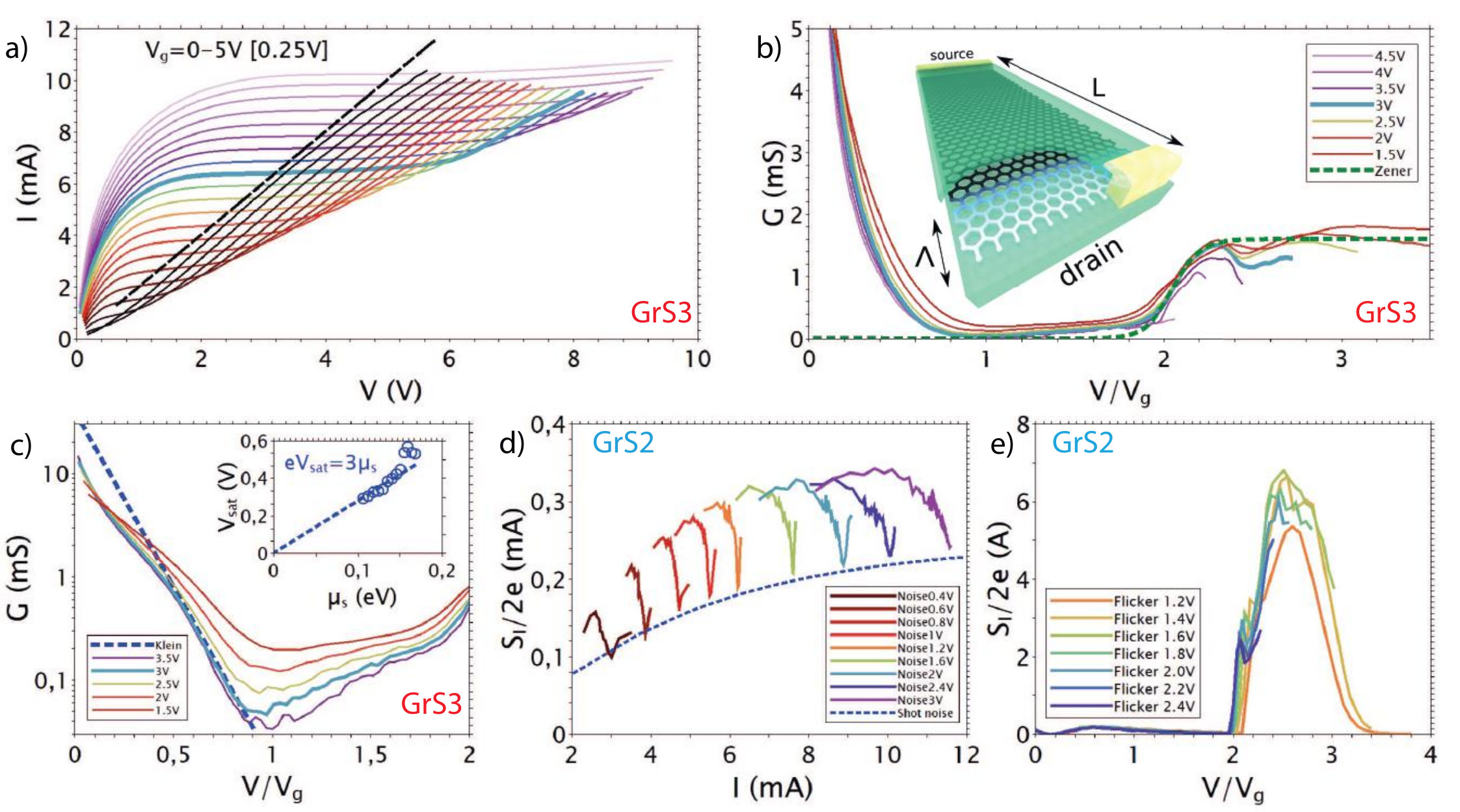}}
\caption{Room temperature ballistic pinchoff in sample GrS3  of dimensions
$L\times W\times t_{hBN}=15\times10\times0.042\;\mathrm{\mu m}$, mobility $\mu=12\mathrm{m^2V^{-1}s^{-1}}$ (gate capacitance $C_g=0.67\mathrm{mF m^{-2}}$), and contact resistance $R_c=120\mathrm{\Omega}$.
Panel-a: Set of current-voltage characteristics after subtraction of contact voltage drop, including the representative $V_{g}=3\;\mathrm{V}$ trace (thick turquoise line).
The pinchoff regime $V=V_{g}$ is indicated by the dashed black line.
Panel-b: differential conductance scaling $G\left[\frac{V}{V_g}\right]$ illustrating the steep decay of saturation conductance below pinchoff,
and the sharp onset of Zener conductance at twice the pinchoff voltage. Inset is an artist view of the supposed carrier density distribution, characterized by a sharp drop over a collimation length $\Lambda$ at the drain side.
Panel-c: semi-log representation of conductance showing the exponential decay at current saturation,
$G=G(0)e^{-\frac{V}{V_{sat}}}$ (dotted blue line), with a doping dependent saturation voltage $V_{sat}\simeq 3\mu_s/e$ (dashed line in the inset).
Panel-d: High-frequency ($1$--$10\;\mathrm{GHz}$) current noise measured in sample GrS2 at $T=10\;\mathrm{K} $  as function of current.
The sharp drop at pinchoff maps the vanishing of the thermal noise contribution $S_I=4 G(I)k_BT_e$, where $T_e$ is the hot-electron temperature, according to the differential conductance dip at saturation. The residual noise (dashed blue line) is attributed to shot noise $S_I=2eI F$ obeying the Fano factor $F\simeq 0.04/\sqrt{1+(I_{sat}/0.006)^2}$. Panel-e shows the huge low frequency ($0.1$-$1\;\mathrm{MHz}$) current noise peak at the onset of the Zener instability.
}
\label{Figure2_Ballistic pinchoff}
\end{figure}

Prior to discussing Schwinger effect in Section \ref{Klein-Schwinger}, it is worth understanding  the electric field profile it stems from, and the current saturation mechanism. We show in Fig.\ref{Figure2_Ballistic pinchoff}-a the high-bias transport characteristics of sample GrS3 (details in the caption of Fig.\ref{Figure2_Ballistic pinchoff}), which exhibit prominent current saturation plateaus centered at the $I(V=V_g)$ pinchoff line (dashed black line). The highlighted  $V_{g}=3\mathrm{V}$ trace (thick turquoise line) illustrates the three main transport regimes: i) the diffusive Drude regime for $V\lesssim V_{sat}$, ii) the Klein-collimation saturation regime for $V_{sat}\lesssim V\lesssim V_Z$, and iii) the Zener regime for $V\gtrsim V_Z$. As discussed in previous works [\onlinecite{Yang2018nnano,Baudin2020adfm}], the Zener effect corresponds to a  field-induced bulk inter-band Klein tunneling and is characterized by a  doping- and bias-independent Zener channel conductivity ($\sigma_Z\sim1\;\mathrm{mS}$). It is subject to Pauli blockade leading up to a finite threshold voltage $V_Z=LE_Z$ with a critical field $E_Z\propto \mu_s^2\propto n_s$ [\onlinecite{Baudin2020adfm}]. Consequently the Zener voltage $V_Z\propto LV_g$ of long transistors is rejected well above $V_g$. This important feature characterizes our device series (see Supplementary Section-IV), which differs from conventional devices  [\onlinecite{Meric2008nnano,Wilmart2020apsci}] where lower mobility and shorter channel lengths substitute saturation plateaus by a crossover between the Drude and Zener ohmic regimes.

In Supplementary Section SI-VI we compare the pinchoff-induced exponential current saturation observed in the present $V_g=Cst$ biasing mode, with the current obtained in the pinchoff-free drain-doping-compensated mode [\onlinecite{Yang2018nnano}]. We also provide a simple 1d-model of the electrostatic potential distribution in the channel to estimate the respective channel and collimation junction voltage drops in the pinchoff regime, thereby establishing the existence of a Klein-collimation junction. Klein collimation is the semi-metal variant of the pinchoff junction of metal-oxide-semiconductor field-effect transistors (MOSFETs) [\onlinecite{Sze2007wiley}]. The exponential saturation is reminiscent of the collimation effect of gate-defined long p-n junctions [\onlinecite{Cheianov2006prbR}], characterized by a transmission $T(k_y)\sim \exp{(-\pi \hbar v_Fk_y^2/eE_x)}$, where $k_y=k_F\sin{\theta}$ is the transverse projection of the Fermi wave vector, $\theta$ the angle of incidence and $E_x=2\mu_s/\Lambda$ the built-in in-plane electric field [\onlinecite{Cheianov2006prbR,Cayssol2009prb,Sonin2009prb}]. This p-n junction model, where electrons with a finite transverse momentum acquire a 1d massive Dirac fermion character with a gap $2\hbar v_F k_y$, provides a natural framework for the bias-induced giant Klein collimation reported here. In contrast to the p-n junction case, Klein collimation entails large $E_x\simeq V/\Lambda$ and $\hbar v_Fk_y=eV\sin\theta_{sat}$, leading to an exponentially vanishing transmission $T(V)\sim \exp{(-\pi V\frac{e\Lambda}{\hbar v_F}\sin^2\theta_{sat})}$. This giant Klein collimation effect turns an incident 2d electron gas into a 1d transmitted beam, leading to an emergent 1d transport regime of relativistic electrons at a velocity $v_F$.

Current saturation is best characterized by the differential conductance $G=\partial I/\partial V$, whose scaling properties as function of $V/V_g$ are shown in Fig.\ref{Figure2_Ballistic pinchoff}-b. The high-bias Zener regime corresponds to a doping-independent step-like increase to $G_Z\simeq1.5\;\mathrm{mS}$ at $V_Z\simeq1.8 V_g$. The saturation regime of interest in this work is characterized by a vanishing conductance in the $[V_{sat},V_Z]$ window. Fig.\ref{Figure2_Ballistic pinchoff}-c is a semi-log plot showing the exponential decay of the saturation conductance above the saturation voltage  $V_{sat}$. It is exemplified by the dashed-blue-line fit of the representative $V_g=3V$ data. This exponential decay is observed over a broad doping range, and characterized by a doping-dependent saturation voltage $V_{sat}$ according to the fitting law  $G=G(0)e^{-V/V_{sat}}$. The inset shows the Fermi-energy dependence of $V_{sat}$ which obeys a linear law $V_{sat}\simeq3\mu_s/e$, implying  that tunneling-conductance suppression at pinchoff, $G(V_g\propto\mu_s^2)\propto e^{-\mu_s/Cte}$, increases with doping. This property is consistently observed in the full sample series (table SI-1) where $eV_{sat}/\mu_s=2.5$--$5$. In the above p-n junction model, the saturation voltage  $V_{sat}=\hbar v_F/\pi e \Lambda\sin^2\theta_{sat}$ is related to the collimation angle $\theta_{sat}$.

An additional signature of the existence of a Klein junction is found in the transistor shot noise in Fig.\ref{Figure2_Ballistic pinchoff}-d. Shot noise $S_I=2e I \mathcal{F}$, where $\mathcal{F}\lesssim 1$ is the Fano factor, is deduced from the microwave current noise $S_I$, which adds to the thermal noise $S_I=4G(I)k_BT_e$ where $T_e$  is the electronic temperature. Here we take advantage of the conductance suppression at saturation, with $G(I\simeq I_{sat})\rightarrow0$ corresponding to the noise dips at the saturation current $I_{sat}$ in Fig.\ref{Figure2_Ballistic pinchoff}-d, to extract the junction Fano factor $\mathcal{F}(I_{sat})\lesssim 0.04$ (see Supplementary Section-III). The presence of a shot noise, and the tiny value of the Fano factor, are strong indications of the presence of a ballistic junction [\onlinecite{Danneau2008prl}]. Finally, the Zener threshold is reached upon increasing further the bias voltage; it entails an upheaval of the electric field distribution, as the electric field, initially confined over $\Lambda$  in the Klein collimation regime, penetrates the full channel length $L$  in the ohmic Zener regime. A signature of this collective instability of Klein collimation can be seen in the huge low-frequency current noise observed over the $[V_Z,1.5V_Z]$ range in Fig.\ref{Figure2_Ballistic pinchoff}-e.

The exponential current saturation, the presence of shot noise, and the Zener electric-field instability are three independent signatures of the giant Klein collimation effect and its collective instability  toward a  bulk Zener effect. The Klein collimation and its associated high-field region  will act as a seed for the Schwinger effect demonstrated below.

\section{Evidence of a universal 1d-Schwinger conductance}\label{Klein-Schwinger}

In order to highlight the small Schwinger-pair contribution in transport, we magnify in  Fig.\ref{Figure3_Klein-Schwinger-effect}-a the conductance-scaling plot of Fig.\ref{Figure2_Ballistic pinchoff}-b. We focus first on the  $V_{g}=3\mathrm{V}$ data (turquoise circles), for which the exponential tails of the Klein collimation (dotted blue line) and Zener (dotted green line) contributions are fully suppressed in a bias window  $[1.1V_g,1.7V_g]$. In this window, the measured conductance is solely governed by the Schwinger contribution, as evidenced by the fit of $V_{g}=3\mathrm{V}$-data with Eq.(\ref{1d-Schwinger-conductance}) taking $V_S=1.4 V_{g}$ (thick turquoise line). Schwinger effect is also visible at lower doping, albeit partially obscured by a residual Klein tunneling contribution. In order to extract the Schwinger-contribution at arbitrary doping, we rely on the de-embedding of the  Klein-tunneling contribution using a three-parameter fit $G=G_S(V/V_S)+G(0)e^{-V/V_{sat}}$ performed in the relevant range $[V_{sat},V_Z]$. The resulting pair conductance $G_S$ is displayed in Fig.\ref{Figure3_Klein-Schwinger-effect}-b. It highlights the sharp transition (dotted black line) separating the Schwinger and Zener regimes, and reveals the Schwinger-conductance fan-like scaling predicted by the affine-approximation  (\ref{1d-Schwinger-conductance-dl}) for a doping-dependent $V_S$.  Remarkably, the linear zero-bias extrapolate is a constant $G_0\simeq0.18\pm0.02\;\mathrm{mS}$, which is very close to the universal quantization $G_0=1.20\frac{4e^2}{h}=0.186\;\mathrm{mS}$ for the 1d-Schwinger effect in graphene. The same procedure has been applied to the full sample series in Fig.SI-(6,7), which shows similar scaling with consistent values of $G_0$. This observation of a robust conductance quantization, in quantitative agreement with 1d-Schwinger theory, is a parameter-free demonstration of the relevance of 1d-Schwinger effect, and of its ubiquity in high-mobility graphene transistors, which constitutes the main finding of our work.

\section{Interpretation of the Schwinger voltage}\label{Schwinger-voltage}

Going one step further, we analyze the dependence of the Schwinger-voltage $V_S(n_s,t_{hBN})$ on doping $n_s$ and hBN-thickness $t_{hBN}$, extracted from theoretical fits of pair-conductance data  with expression (\ref{1d-Schwinger-conductance}).
We assume that $V_S(n_s,t_{hBN})=E_S(n_s)\Lambda(n_s,t_{hBN})$ with $E_S=\Delta_S^2(n_s)/e\hbar v_F$ for a Schwinger gap $\Delta_S$ set by a collimation gap  $2\hbar v_F k_S$ at finite transverse momentum $k_S$. In high-energy physics, the Schwinger threshold for electron-positron creation in vacuum (absence of any other particles before pair creation) is set by  a true spectral gap $2 m c^2$. In condensed matter, the threshold for electron-hole creation can also be associated to a Pauli blocking effect related to the presence of other electrons filling graphene bands before pair creation. For the Klein-Schwinger effect we assume heuristically that this threshold energy
 is set by $2\hbar v_F k_s$, where $k_s$ is a typical transverse momentum for the fully populated Klein transmitted electrons. In the absence of a full-fledged theory, and for the purpose of estimating the Klein-collimation length $\Lambda(n_s,t_{hBN})$, we use below the Ansatz $\Delta_S\simeq\mu_s$.

\begin{figure}[h!!!!]
\centerline{\includegraphics[width=16cm]{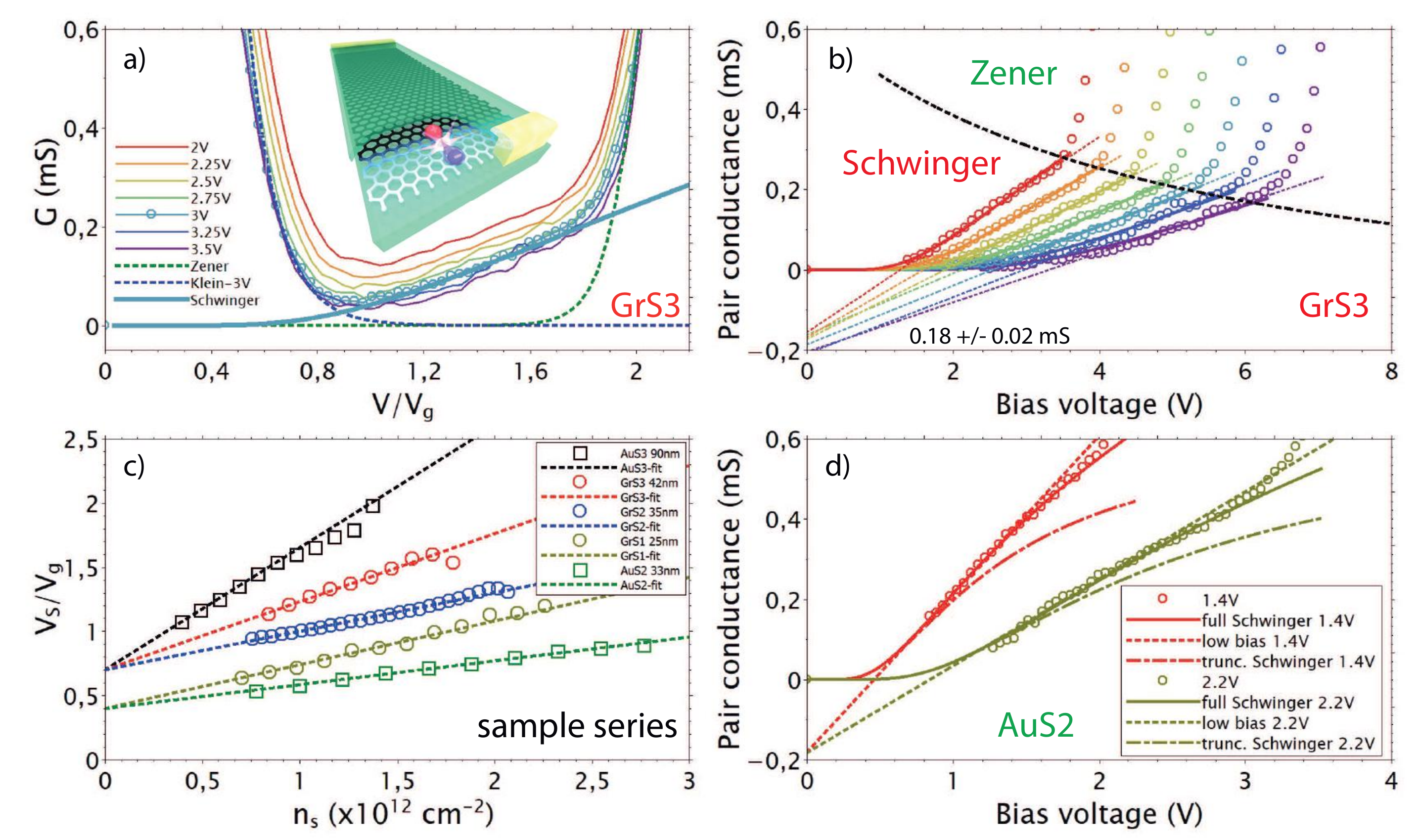}}
\caption{Klein-Schwinger effect in long ballistic transistors.
Panel-a: blow-up of GrS3 conductance in Fig.\ref{Figure2_Ballistic pinchoff}-b. The fully developed gap at large doping (dashed blue and green lines of the $V_{g}=3\;\mathrm{V}$ data), unveils the Schwinger-pair contribution (thick turquoise line) corresponding to $V_S\simeq 1.4 V_{g}$. Inset: artist view of the Klein-Schwinger mechanism. At lower doping, where the Schwinger contribution is embedded in a finite Klein contribution, the pair contribution is extracted from the single particle tunneling contribution by a 3-parameters fit of conductance data according to the law $G(V)=G_S(V/V_S)+G(0)e^{-V/V_{sat}}$.
Panel-b: the pair contribution $G(V)-G(0)e^{-V/V_{sat}}$ (same color code as in panel a), reveals the affine-approximation scaling of the Schwinger conductance (\ref{1d-Schwinger-conductance-dl}) with $G_0\simeq0.18\pm0.02\;\mathrm{mS}$ and  $V_S=0.7V_g+0.15V_g^2$. The Zener regime is signaled by a sharp increase of pair conductance above the Schwinger-Zener boundary (dotted black line) set by $G_{S}(V=V_Z=1.9V_g)$.
Panel-c: linear increase of the $V_S/V_g(n_s)$ ratio with doping observed in the full transistor series (see Supplementary Fig.SI-7) where  Schwinger-pair conductance scaling is reported with consistent values of $G_0$. The slope increases with dielectric thickness (see main text). The $V_S/V_g$ ratio is minimized in sample AuS2 ($L\times W\times t_{hBN}=10.5\times15\times0.035\;\mathrm{\mu m}$) and maximized in sample AuS3 ($L\times W\times t_{hBN}=11.1\times11.4\times0.090\;\mathrm{\mu m}$).
Panel-d: broad range ($V/V_S\lesssim4$) Schwinger conductance at low doping in AuS2 where $V_S/V_g\simeq0.5$ showing the quantitative agreement with Eq.(\ref{1d-Schwinger-conductance}) (solid lines) that deviates from the affine-approximation Eq.(\ref{1d-Schwinger-conductance-dl}) approximation (dotted lines), and from the truncated approximation $G_S^*$  (dash-dotted lines).
}
\label{Figure3_Klein-Schwinger-effect}
\end{figure}

We plot in Fig.\ref{Figure3_Klein-Schwinger-effect}-c the ratio $V_S/V_g(n_s)$ for the full device series. We observe a linear doping dependence with a slope that increases with gate dielectric thickness $t_{hBN}=25$--$90\mathrm{nm}$. As detailed in the Supplementary Section-V, these dependencies can be explained by $\Lambda(n_s,t_{hBN})$, according to the law  $\frac{\Lambda}{t_{hBN}}\frac{\Delta_S^2}{\mu_s^2}=4\alpha_g\frac{V_S}{V_g}$, where
$\alpha_g=\frac{e^2}{4\pi\epsilon_{hBN}\varepsilon_0\hbar v_F}=0.70$ (with $\epsilon_{hBN}=3.1$ at high field  [\onlinecite{Pierret2022MatRes}]) is the graphene fine structure constant. Assuming that $\Delta_S\simeq\mu_s$, we can cast the measured $\frac{V_S}{V_g}(t_{hBN},n_s)$ into the $t_{hBN}$ power expansion $\Lambda\simeq a t_{hBN}+\xi n_st_{hBN}^2$, where $a=1$--$2$ is a geometrical factor depending on the details of the contact-gate arrangements, and $\xi\simeq 4\;\mathrm{nm}$ is a microscopic interaction length \footnote{$\xi$ is the typical interaction radius per electron (for $n_s t_{hBN}^2=1$) fitting in a junction length $\Lambda\sim t_{hBN}$.}, quantifying the doping-induced dilation of the junction length. This Coulomb-repulsion effect enhances $V_S$, favoring the visibility of the Klein-Schwinger effect at large doping (GrS3-data in Fig.\ref{Figure3_Klein-Schwinger-effect}-c), up to the limit where its onset exceeds $V_Z$ so that Schwinger effect becomes masked by the Zener instability (AuS3-data in Fig.\ref{Figure3_Klein-Schwinger-effect}-c). The latter case prevails in thick-hBN samples, giving rise to  extremely flat current-saturation plateaus (see Fig.SI4-f). Conversely, the small $V_S/V_g\simeq0.5$ ratio of thin-hBN samples (AuS2-data in Fig.\ref{Figure3_Klein-Schwinger-effect}-c),  allows investigating conductance over an extended bias range $V/V_S\lesssim4$. The doping and hBN-thickness dependencies of the ratio $V_S/V_g$, which contrast with the constant $V_Z/V_g$, support our interpretation  of  Schwinger effect as the intrinsic breakdown mechanism of Klein collimation, rather than a precursor of the extrinsic (length dependent) collective Zener instability.

We conclude our experimental report by analyzing the extended Schwinger-conductance regime in the AuS2-data which are plotted in Fig.\ref{Figure3_Klein-Schwinger-effect}-d. The accessible experimental range of Schwinger conductance ($G_S\lesssim0.6\;\mathrm{mS}$) exceeds the validity domain ($G_S\lesssim0.2\;\mathrm{mS}$) of the affine approximation (\ref{1d-Schwinger-conductance-dl}), unveiling the sub-linearities involved in Eq.(\ref{1d-Schwinger-conductance}) (solid lines). Experimental data are in excellent agreement with the full 1d-Schwinger prediction, significantly deviating from the affine  approximation (dotted black line), and strongly deviating from the truncated-rate 1d-Schwinger conductance $G_S^*$ (dash-dotted lines). This observation constitutes a complementary evidence of the relevance of the Schwinger theory [\onlinecite{Schwinger1951prb}].

The demonstration of the Schwinger effect in an effective field theoretical 1+1 dimensional system is the first of its kind and a confirmation of a crucial prediction of quantum electrodynamic field theory. It fulfills the promise of using graphene to emulate quantum electrodynamics [\onlinecite{Novoselov2011rmp}], specifically here in its strong field sector. The use of condensed matter analogs has already proven fruitful in cosmology [\onlinecite{Zurek1996prep}], in particular with the observation of vortex formation in neutron-irradiated superfluid He-3 as an analogue of cosmological defect formation [\onlinecite{Ruutu1996nature}], or that of the analogue of black hole Hawking radiation [\onlinecite{deNova2019nature}], as well as in the understanding of energy renormalization via Lorentz boosts [\onlinecite{Wyzula2022advsci}].

Our experiment shows the  ability of giant Klein collimation to generate large local electric fields. This opens a way for the exploration of Schwinger effect in different systems, like the massive Dirac fermions in bilayer graphene [\onlinecite{Katsnelson2012jetp}], or the massless fermions in 3d Weyl or Dirac semi-metals [\onlinecite{Katsnelson2014jltp}].  In this respect, theoretical challenges remain concerning the modeling of strongly out-of-equilibrium collimation and the emergence of the Schwinger gap. Our experiment also shows that prominent current saturations, with large saturation velocities, can be obtained in gapless graphene. On the application side, the understanding of the Klein-Schwinger mechanism turns out to be the key for the optimization of large voltage-gain $A =\partial V/\partial V_g=G_m/G$ (see transconductance $G_m$ in Fig.SI-2)  in high-mobility graphene transistors, which is tunable from $A\sim10$ in thin-hBN AuS2 to $A\sim100$ in thick-hBN AuS3 according to the $V_S/V_g$ ratio (see Table.SI-1). Our work thus describes the implementation of a fully-relativistic Klein-Schwinger field-effect transistor showing large saturation currents and voltage gain.

Finally one may attempt to go deeper in the condensed matter analogy of QED, by investigating other manifestations of Schwinger effect like the full counting statistics of pair creation, or the vacuum polarization [\onlinecite{Schwinger1951prb}] which is a precursor of pair creation. This raises the question of the  dynamics of Schwinger-pair creation, which is an open field that can be investigated by dynamical transport and electromagnetic radiation spectroscopy.

\section{Acknowledgments}

The research leading to these results has received partial funding from the European Union Horizon 2020 research and innovation program under grant agreement No.881603 Graphene Core 3 (DM, MR, EBa, BP).

\section{Authors contributions statement}

AS, EBa and BP conceived the experiment. AS conducted device fabrication and measurements, under the guidance of DM and MR in the early developments. TT and KW have provided the hBN crystals. PV, JC, MOG, EBa, JT and BP developed the models and theoretical interpretations. AS, DM, CV, JMB, EBo, GF, CV, EBa and BP participated to the data analysis. AS, EBa, and BP wrote the manuscript with contributions from the coauthors.

\section{Competing Interests Statement}

The authors declare no competing interests.

\newpage

\section{Figure Captions}

\section{Methods}

\textit{Fabrication of ballistic graphene transistors.} The hexagonal boron nitride encapsulated graphene heterostructures are fabricated with the standard pick-up and stamping technique, using a polydimethylsiloxane(PDMS)/ polypropylenecarbonate (PPC) stamp [\onlinecite{Yankowitz2019Nature}]. The gate is first fabricated on a high-resistivity Si substrate covered by 285nm $\text{SiO}_{\text{2}}$. Gate electrode is either a thin exfoliated graphite flake (thickness $\lesssim 15nm$) or a pre-patterned gold pad (thickness $70nm$) designed by laser lithography and Cr/Au metallisation. Deposition of the heterostructure on the backgate is followed by acetone cleaning of the stamp residues, Raman spatial mapping and AFM characterization of the stack. Graphene edge contacts are then defined by means of laser lithography and reactive ion etching, securing low contact resistance $\lesssim 1\;\mathrm{k\Omega\mu m}$. Finally, metallic contacts to the graphene channel are designed with a Cr/Au Joule evaporation, that also embeds the transistor in a coplanar waveguide geometry suited for cryogenic probe station microwave and noise characterization.
The large transistor dimensions $L,W \gtrsim 10 \mathrm{\mu m} $ secure moderate channel electric field $E\sim V/L \lesssim 10^{6} \mathrm{V/m}$ up to the MOSFET pinchoff at $V=V_{g}$, while their high mobility $\mu \gtrsim 6\mathrm{m^{2}V^{-1}s^{-1}}$ at room temperature, and $\mu \gtrsim 35 \mathrm{m^{2}V^{-1}s^{-1}} $ at 10K, secures ballistic transport in the channel.

\textit{RF transport and noise measurement.} Characterization of the ballistic graphene transistors is performed in a cryogenic probe station adapted to RF measurements up to 67GHz. The DC measurements are performed using a Keithley 2612 voltage source to apply gate and bias. Noise measurements are enabled by the use of a Tektronix DPO71604C ultrafast oscilloscope for measurements up to $16 $GHz. The high-frequency signal coming from the device is amplified by a CITCRYO1-12D Caltech low noise amplifier in the $1-10$ GHz band, whose noise and gain have been calibrated against the thermal noise of a $50 \Omega$ calibration resistance measured at various temperatures between 10K and 300K in the probe station. For low frequency noise measurements in the $0.1-1$ MHz band, a NF-Corporation amplifier (SA-220F5) is used.

\section{Data availability}

Data are available at the DOI : https://www.doi.org/10.5281/zenodo.7104630

\end{document}


\title{Mesoscopic Klein-Schwinger effect in  graphene \\ (Supplementary Information)}

\author{A. Schmitt}\email{aurelien.schmitt@phys.ens.fr}
\affiliation{Laboratoire de Physique de l'Ecole normale sup\'erieure, ENS, Universit\'e
PSL, CNRS, Sorbonne Universit\'e, Universit\'e de Paris, 24 rue Lhomond, 75005 Paris, France}
\author{P. Vallet}
\affiliation{Laboratoire Ondes et Mati\`ere d’Aquitaine, 351 cours de la lib\'eration, 33405 Talence, France}
\author{D. Mele}
\affiliation{Laboratoire de Physique de l'Ecole normale sup\'erieure, ENS, Universit\'e
PSL, CNRS, Sorbonne Universit\'e, Universit\'e de Paris, 24 rue Lhomond, 75005 Paris, France}
\affiliation{Univ. Lille, CNRS, Centrale Lille, Univ. Polytechnique Hauts-de-France, Junia-ISEN, UMR 8520-IEMN, F-59000 Lille, France.}
\author{M. Rosticher}
\affiliation{Laboratoire de Physique de l'Ecole normale sup\'erieure, ENS, Universit\'e
PSL, CNRS, Sorbonne Universit\'e, Universit\'e de Paris, 24 rue Lhomond, 75005 Paris, France}
\author{T. Taniguchi}
\affiliation{Advanced Materials Laboratory, National Institute for Materials Science, Tsukuba,
Ibaraki 305-0047,  Japan}
\author{K. Watanabe}
\affiliation{Advanced Materials Laboratory, National Institute for Materials Science, Tsukuba,
Ibaraki 305-0047, Japan}
\author{E. Bocquillon}
\affiliation{Laboratoire de Physique de l'Ecole normale sup\'erieure, ENS, Universit\'e
PSL, CNRS, Sorbonne Universit\'e, Universit\'e de Paris, 24 rue Lhomond, 75005 Paris, France}
\affiliation{II. Physikalisches Institut, Universit\"at zu K\"oln, Z\"ulpicher Strasse 77, 50937 K\"oln}
\author{G. F\`eve}
\affiliation{Laboratoire de Physique de l'Ecole normale sup\'erieure, ENS, Universit\'e
PSL, CNRS, Sorbonne Universit\'e, Universit\'e de Paris, 24 rue Lhomond, 75005 Paris, France}
\author{J.M. Berroir}
\affiliation{Laboratoire de Physique de l'Ecole normale sup\'erieure, ENS, Universit\'e
PSL, CNRS, Sorbonne Universit\'e, Universit\'e de Paris, 24 rue Lhomond, 75005 Paris, France}
\author{C. Voisin}
\affiliation{Laboratoire de Physique de l'Ecole normale sup\'erieure, ENS, Universit\'e
PSL, CNRS, Sorbonne Universit\'e, Universit\'e de Paris, 24 rue Lhomond, 75005 Paris, France}
\author{J. Cayssol}
\affiliation{Laboratoire Ondes et Mati\`ere d’Aquitaine, 351 cours de la lib\'eration, 33405 Talence, France}
\author{M. O. Goerbig}
\affiliation{Laboratoire de Physique des Solides, CNRS UMR 8502, Univ. Paris-Sud, Universit\'e
Paris-Saclay, F-91405 Orsay Cedex, France}
\author{J. Troost}
\affiliation{Laboratoire de Physique de l'Ecole normale sup\'erieure, ENS, Universit\'e
PSL, CNRS, Sorbonne Universit\'e, Universit\'e de Paris, 24 rue Lhomond, 75005 Paris, France}
\author{E. Baudin} \email{emmanuel.baudin@phys.ens.fr}
\affiliation{Laboratoire de Physique de l'Ecole normale sup\'erieure, ENS, Universit\'e
PSL, CNRS, Sorbonne Universit\'e, Universit\'e de Paris, 24 rue Lhomond, 75005 Paris, France}
\author{B. Pla\c{c}ais} \email{bernard.placais@phys.ens.fr}
\affiliation{Laboratoire de Physique de l'Ecole normale sup\'erieure, ENS, Universit\'e
PSL, CNRS, Sorbonne Universit\'e, Universit\'e de Paris, 24 rue Lhomond, 75005 Paris, France}




\maketitle

\tableofcontents
\newpage 

\renewcommand{\thefigure}{SI-\arabic{figure}}

\section{Schwinger effect in various dimensions}

Quantum electrodynamics predicts a pair creation rate, per unit volume, area or length depending on dimensionality, as given by [\onlinecite{Zuber2006dover}], with $E_{S}=\frac{m^2c^3}{e\hbar}$ the Schwinger field,
\begin{equation}
w_{3d}=\frac{(eE)^2}{4\pi^3c\hbar^2}\sum_{n\geq1}\frac{e^{-n\pi \frac{E_S}{E}}}{n^{2}}\quad,\quad
w_{2d}=\frac{eE}{2\pi^2\hbar}\sqrt{\frac{eE}{c\hbar}}\sum_{n\geq1}
\frac{e^{-n\pi \frac{E_S}{E}}}{n^{3/2}}\quad,\quad
w_{1d}=\frac{eE}{\pi\hbar}\sum_{n\geq1}\frac{e^{-n\pi \frac{E_S}{E}}}{n}\quad.
\label{QED-Schwinger} \end{equation}

When adapted to condensed matter, speed of light $c$ is replaced by Fermi velocity $v_{F}$ and the rest energy is substituted with the gap $\Delta_S$. The pair partners being dissociated in the strong electric field, one can thus write a mesoscopic pair current, with a mesoscopic Schwinger field $ E_S=\frac{\Delta_S^2}{e\hbar v_F}$ and voltage $V_{S}=\frac{\Lambda\Delta_S^2}{e\hbar v_F}$, as
\begin{equation}
J_{3d}=\frac{e^2E^2\Lambda}{2\pi^3v_F\hbar^2}\sum_{n\geq1}
\frac{e^{-n\pi \frac{E_S}{E}}}{n^{2}} \quad , \quad
J_{2d}=\frac{2g_sg_ve^2}{\pi h}\sqrt{\frac{e\Lambda^2}{\hbar v_F}}E^{3/2}\sum_{n\geq1}
\frac{e^{-n\pi \frac{E_S}{E}}}{n^{3/2}} \quad
\label{Meso-Schwinger} \end{equation}

\begin{equation}
I_{1d}=\frac{4g_sg_ve^2}{h} V_{ds}\sum_{n\geq1}\frac{e^{-n\pi \frac{V_S}{V_{ds}}}}{n}\quad.
\label{Meso-Schwinger2} \end{equation}

Not specific to Schwinger effect, the  $J_{3d}\propto E^{2}e^{-\frac{Cte}{E}}$ is reminiscent of
field emission according to the Fowler-Nordheim mechanism,
which is a non-relativistic variant, corresponding to quantum tunneling across a triangular barrier  [\onlinecite{Sze2007wiley,Grado2007optik}], with
\begin{equation}
J_{FN}=\frac{e^3m_0E^2}{16\pi^2m^*\hbar\phi} e^{-\pi\frac{E_{FN}}{E}}\qquad,\qquad E_{FN}=\frac{4\sqrt{2m^*\phi^3}}{3\pi e\hbar}\sim \frac{4\sqrt{2}}{3\pi}\frac{\phi^2}{e\hbar v_F}\quad,
\label{Fowler-Nordheim}\end{equation}
where $m_0$ and $m^*$ are the bare and effective electron masses,  $\phi$ is the barrier height.  It differs from Eq.(\ref{Meso-Schwinger})
in  the non-universal character of the pre-factor and critical field.

Mesoscopic Schwinger effect has been considered to describe non-linear transport in gapless neutral 2d-graphene [\onlinecite{Gavrilov1996prd,Shytov2009ssc,Dora2010prb,Vandecasteele2010prb,Gavrilov2012prd}],
 with :
\begin{equation}
E_{S}=0\qquad J_{2d}=2.612\frac{2g_sg_ve^2}{\pi h}\sqrt{\frac{e\Lambda^2}{\hbar v_F}}E^{3/2}\quad,
\qquad\label{2d-Schwinger} \end{equation}
where $\Lambda$ is the length of the charge neutrality region, and $g_s=g_v=2$ are the spin and valley degeneracy of graphene.
The non-linear Schwinger-pair contribution exceeds the single particle tunneling conductivity $\sigma_{2d}=\frac{g_sg_ve^2}{\pi h}$ [\onlinecite{Tworzydlo2006_prl,Danneau2008_prl}]
above a bias voltage $V_{2d}\sim\hbar v_F/e\Lambda\sim 1\mathrm{mV}$ (for $\Lambda=1\mathrm{\mu m}$).

\begin{figure}[h!]
\hspace*{-1cm}
\centering{}\includegraphics[scale=0.5]{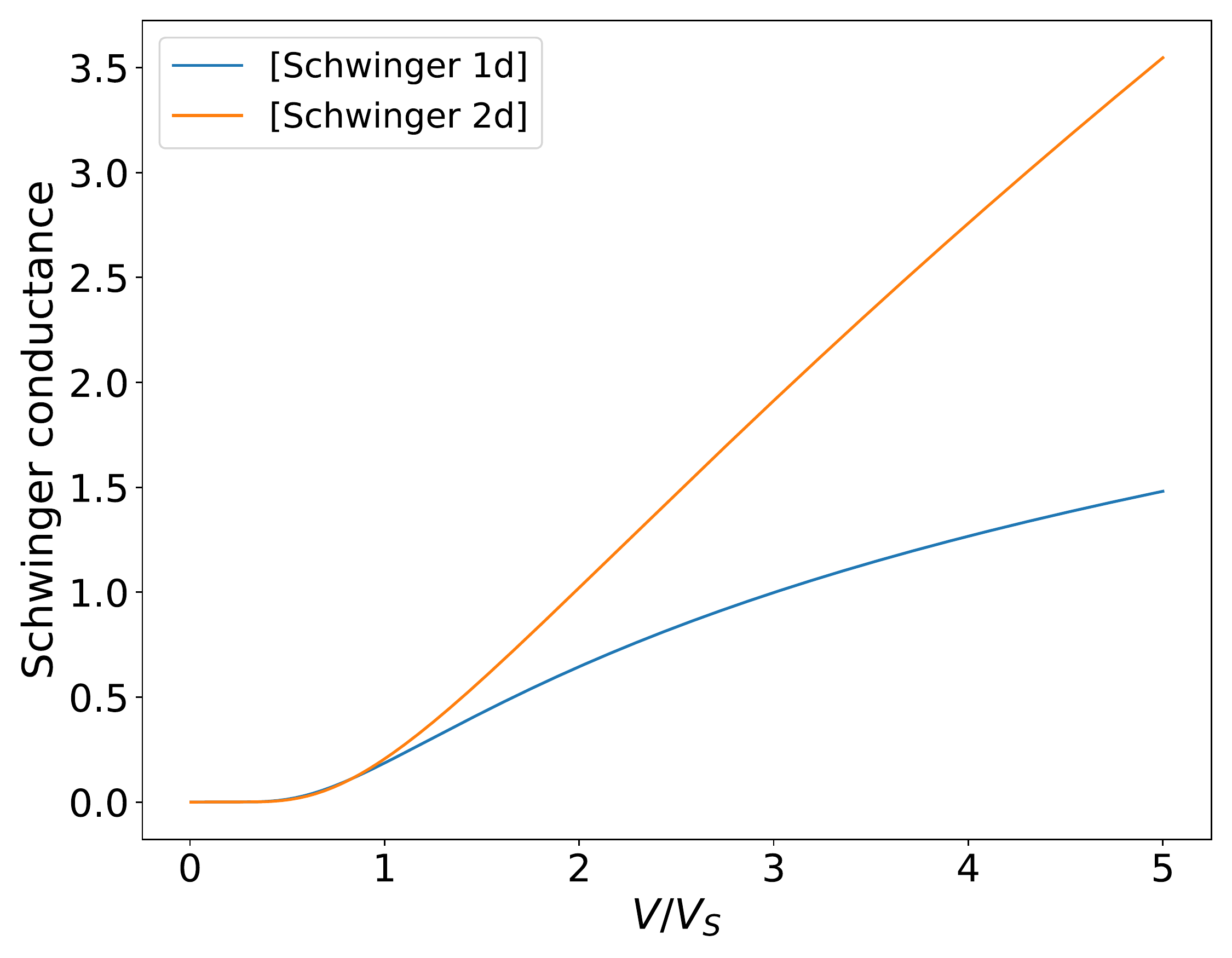} \caption{Representation of the functional forms (terms in brackets) of equations (\ref{conductance-1d}) and (\ref{conductance-2d}) for 1d and 2d massive Schwinger effect.
The two forms have the same threshold and low-bias development but differ at large bias. The  2d conductance is not universal and depends on the ratio of the sample width $W$ to the Compton length  $\Lambda_C$ which is typically a large number. It is larger but also linear over a wider bias range than the 1d-Schwinger conductance.}
\label{1d2d}
\end{figure}

The 2d massive Schwinger effect differs from the 1d massive Schwinger effect studied in the main text. The 1d and 2d conductance read : 
\begin{eqnarray}
    G_{1d}=4\;\;\; \bigg( \frac{g_{s}g_{v}e^{2}}{h} \bigg)&\times& \Bigg[ \pi \sqrt{\frac{V}{V_{S}}}\;\frac{1}{e^{\pi V_{S}/V}-1}\;+\; \mathrm{ln}\left(\frac{1}{1-e^{-\pi V_{S}/V}}\right) \Bigg]\label{conductance-1d}\\
    G_{2d}=\frac{2W}{\pi \Lambda_{C}} \bigg( \frac{g_{s}g_{v}e^{2}}{h} \bigg)&\times& \Bigg[ \frac{3}{2} \sqrt{\frac{V}{V_{S}}} \sum_{n \geq 1} \frac{e^{-n\pi V_{S}/V}}{n^{3/2}} + \pi \sqrt{\frac{V_{S}}{V}} \sum_{n \geq 1} \frac{e^{-n\pi V_{S}/V}}{n^{1/2}} \Bigg]\quad ,\label{conductance-2d}
\end{eqnarray}
where $W$ is the sample width and  $\Lambda_{C}=\sqrt{\frac{v_{F} \hbar}{e E_{S}}}\lesssim10\mathrm{nm}$ the Compton length. Compared to the 1d Schwinger conductance, the 2d-conductance prefactor is  not universal and much larger by a factor $W/\Lambda_{C}(V_g)\sim1000$. Evidence of the universal character of the measured conductance in the ballistic transistors is given in Figure \ref{Pair6plots} and Figure 3-b from the main text, that all show the same universal zero-bias affine extrapolate $G_0=0.18\pm0.02 ~ \mathrm{mS}$ despite contrasted values of $W$ and $E_{S}$. Apart from the prefactor, the two formulae also show different behavior with respect to bias voltage, as shown in Figure \ref{1d2d}. The sublinear dependence of the 1d-conductance of Eq.(\ref{conductance-1d}) is clearly seen in Fig. \ref{Pair6plots}d) and e), and in Figure 3-d from the main text.

\begin{figure}[h!]
\hspace*{-1cm}
\centering{}\includegraphics[scale=0.3]{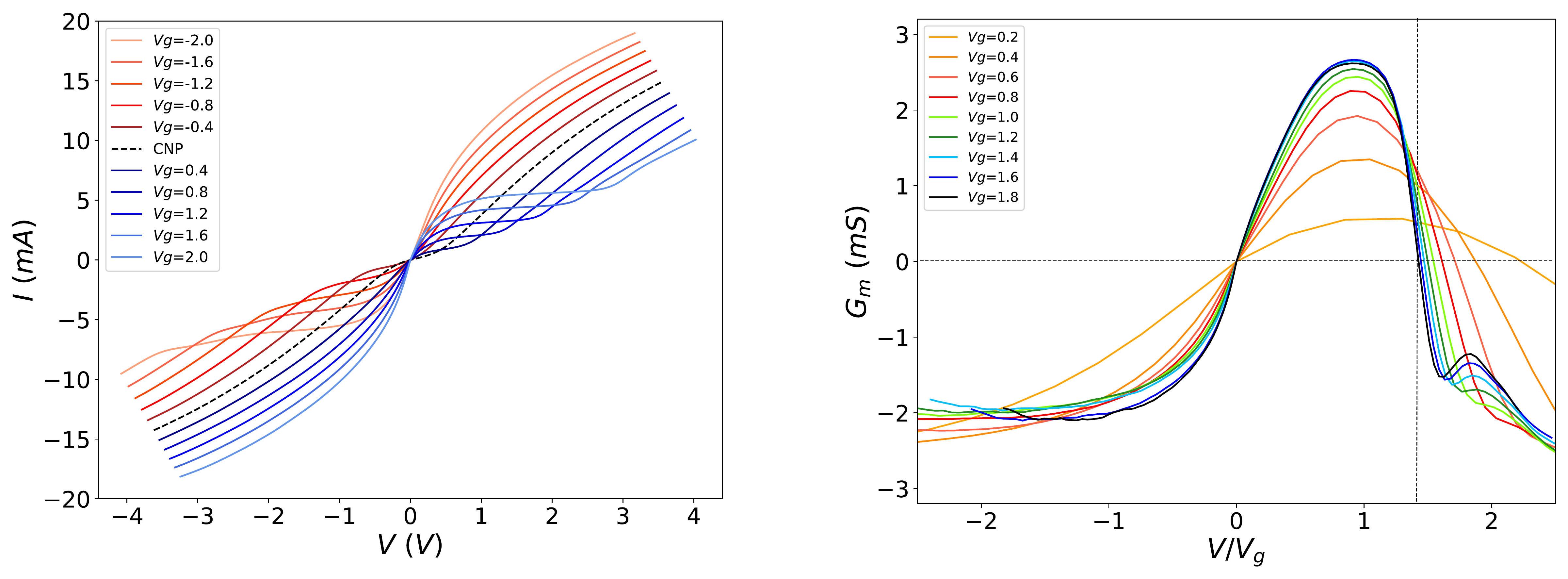} \caption{ Overview of ballistic pinch-off in hBN-encapsulated graphene transistor GrS1, of
dimensions $L\times W\times t_{hBN}=13\times17\times0.025\;\mathrm{\mu m}$,
mobility $\mu=6.3\ \mathrm{m^2V^{-1}s^{-1}}$, and contact resistance $R_c=80\;\mathrm{\Omega}$. Left: full bipolar representation. Pinch-off free transport is observed in the  drain carrier accumulation regime for $\mathrm{Sign}(V\times V_{g})<0$.
It shows standard intra-band velocity saturation by optical phonons (OP), followed by the Zener regime.
Pinch-off is observed under drain depletion, for a channel voltage $V=V_{bias}-R_cI\sim V_{g}$,
with broad ($V\sim0.3$--$1.5 V_{g}$) current plateaus.
Pinch-off current is depleted by a factor two below the drain-accumulation counterpart,
falling below the massless charge neutrality Zener level (CNP, black dotted line).
Pinch-off plateaus terminate as an instability toward the ohmic Zener regime, which is rejected here above $V_{Z}\gtrsim 1.5 V_{g}$ by Pauli blocking. Right: DC transconductance scaling as function of $V/V_{g}$, concentrating on positive $V_{g}$ values. It underlines the asymmetry between positive and negative $V$ regimes, with a negative transconductance for negative $V$, illustrating the increase of current with increasing doping. For positive $V$, the transconductance turns strongly negative, highlighting the depletion of the pinch-off current below the zero-doping CNP value. The transconductance changes sign slightly above pinch-off at $V \simeq 1.4 V_{g}$, corresponding to a current $I \simeq 1.1 I_{sat}$.}
\label{GRS1_overview}
\end{figure}

\section{Overview of ballistic pinch-off in sample GrS1}

A complete panorama of ballistic pinch-off in graphene transistors can be found in sample GrS1, of dimensions $L\times W\times t_{hBN}=13\times17\times0.025\;\mathrm{\mu m}$, shown in Fig.\ref{GRS1_overview}. Thanks to fortuitous equal electron and hole mobilities ($\mu = 6.3\; \mathrm{m}^{2}\mathrm{V}^{-1}\mathrm{s}^{-1}$), the current-voltage characteristics in Fig.\ref{GRS1_overview}-left are fully symmetric. We can distinguish between two different regimes. When $\mathrm{Sign}(V\times V_{g})<0$, increasing $V$ causes an accumulation of charge carriers on the drain side, but the channel stays unipolar. In this case we observe pinch-off-free transport, which shows standard intra-band velocity saturation by optical phonons followed by the Zener inter-band regime with a doping-independent conductance $G_Z(V)\sim 2.5\mathrm{mS}$ [\onlinecite{Yang2018nnano}]. When $\mathrm{Sign}(V\times V_{g})>0$, pinch-off is observed under drain depletion for a voltage $V \sim V_{g}$, with broad current saturation plateaus ($V\sim0.3$--$1.5 V_{g}$). These plateaus terminate as an instability towards the ohmic Zener regime, where charge transport becomes bipolar in the channel. Due to the carrier depletion at the drain, the pinch-off current is depleted below the massless charge neutrality Zener level, suggesting the existence of a doping-induced conductance gap.

The two different transport regimes can also be identified in the transconductance $G_{m}$ in Fig.\ref{GRS1_overview}-right. For sake of visibility, we focus here on the $V_{g} >0$ curves. The drain carrier accumulation for $V<0$ is associated with a negative transconductance, as higher gate values, corresponding to higher doping, lead to higher current. On the other hand, the negative values of $G_{m}$ for positive bias confirm the pinch-off depletion of the current.

Unfortunately, the sample GrS1 suffers from a relatively low electronic mobility and a low dielectric thickness $t_{hBN}$, which results in a early onset of the Zener conductance and a reduced accessible voltage range. The hBN thickness indeed determines the range of accessible bias and gate voltages that can be applied on the device without generating strong leak currents and finally the breakdown of the insulator [\onlinecite{Pierret2022MatRes}] ; the leakage can be due to a high gate voltage, but can also appear below the drain electrode if the voltage difference $V-V_{g}$ becomes too important. As a consequence, the raw differential conductance shows no signature of Schwinger effect. 

However, the subtraction of the fitted Klein conductance allows restoring the visibility of the 1d-Schwinger conductance below the Zener onset voltage, as shown in Figure \ref{Pair6plots}-a. Further study of this device can be found in Section \ref{section:dispos}.

\section{Noise characterization of  Klein collimation}

\begin{figure}[h!]
\hspace*{-1.1cm}
\centering{}\includegraphics[scale=0.19]{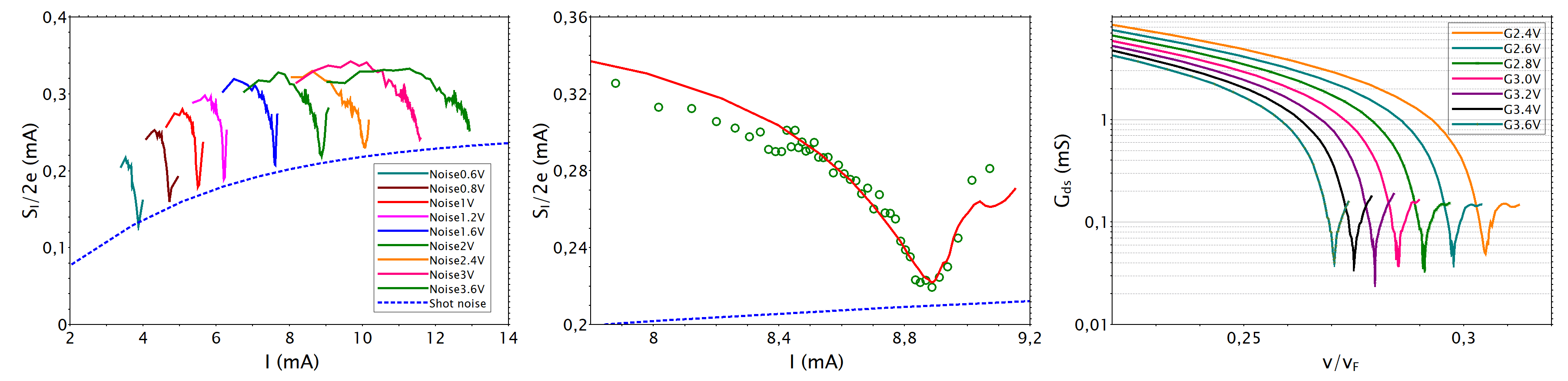} \caption{Current noise signature of Klein-collimation pinch-off in sample GrS2. Dimensions are
$L\times W\times t_{hBN}=10.5\times 15\times 0.035\;\mathrm{\mu m}$.
Left: high-frequency white noise $S_I(I)$,  measured at $T_0=10\;\mathrm{K}$ in the $1$--$10\;\mathrm{GHz}$ band using Caltech Low Noise Amplifiers (model CITCRYO1-12D). It shows
a sharp dip at pinch-off mapping the conductance dip (right panel).
Note that the noise dip occurs at a current 10 percent larger than the conductance dip, due to the transconductance correction to the Johnson-Nyquist formula for a transistor $S_{therm}\simeq 4|G+\beta G_m|k_BT_e$ .
Considering that noise is the sum of a  thermal contribution, $S_{th}=4Gk_BT_e$, and a shot-noise contribution,
$S_{shot}=2e I\mathcal{F}$ where $\mathcal{F}\lesssim1$ is a Fano factor, we extract the pinch-off shot noise
$S_{shot}/2e(I_{sat})\simeq I_0\times I_{sat}/\sqrt{I_1^2+I_{sat}^2}$ (dashed blue line),
with $I_0=0.26\;\mathrm{mA}$ and  $I_1=6\;\mathrm{mA}$. This translates into a tiny pinch-off current Fano factor,
$\mathcal{F}=I_{shot}/I_{sat}\simeq I_0/\sqrt{I_{sat}^2+I_1^2}\lesssim 0.04$,
that vanishes for $I_{sat}\gg I_1$. Center: Zoom on the high-frequency white noise $S_I(I)$ close to pinch-off for $V_{g}=2V$.  Using the measured $G(I)$,
we fit noise data (red line) to estimate the electronic temperature on the current saturation plateau ($I\simeq I_{sat}$)
at $T_e\sim1100\mathrm{K}\times V/V_{g}$  ($V_{g}=2\;\mathrm{V}$), which corresponds to  $k_BT_e/eV_{g}\simeq0.1$.
Right: Conductance-velocity $G(v/v_{F})$  representation of ballistic pinch-off at $T_{0}=10\;\mathrm{K}$. The velocity is defined as the velocity at the source electrode by $v=\frac{I}{C_{g}V_{g}W}$.In this representation,
the broad $V\simeq 1$--$2V_{g}$ conductance gap collapses into a singular point at $G(I_{sat})\lesssim0.1\;\mathrm{mS}$.}
\label{bruit}
\end{figure}

Additional signatures of ballistic pinch-off can be found in the current noise $S_I$, which is measured in sample GrS2. It is representative of the series and actually quite similar to sample GrS3 of the main text. We distinguish different noise contributions according to frequency: the low-frequency flicker noise, which is measured at sub-MHz frequency and is described in the main text,
and the  flicker-free white thermal/shot noise, which is measured at GHz frequency.

The GHz noise (measured at $T=10\;\mathrm{K}$) is plotted in Fig.\ref{bruit}-left as function of current.
It exhibits a sharp dip at current saturation ($I\gtrsim I_{sat}$) which is reminiscent of the conductance dip observed in the $G(v/v_{F})$ data of Fig.\ref{bruit}-right.
We thus analyze the $S_I(I)$ dependence as the sum of two terms: a thermal contribution $S_{therm}\simeq 4G_{ds}k_BT_e$,
where $T_e$ is the hot-electron temperature, and a shot-noise contribution $S_{shot}=2e I\mathcal{F}$,
where $\mathcal{F}\lesssim1$ is a Fano factor. 
In this interpretation, the GHz noise dip corresponds to the vanishing of
$S_{therm}(I_{sat})\propto G(I_{sat})\rightarrow0$ at current saturation, leaving shot noise as the residual noise.
This assumption is supported by the $S_{I}(I)-S_{I}(I_{sat})\propto S_{therm}(I)$ dependence which can be fitted in the current
saturation regime assuming a  $T_e(I)=T_e(I_{sat})V/V_{g}$ proportional to Joule power $P\simeq I_{sat}V$ (red line in
Fig.\ref{bruit}-center with $T_e\sim1100\mathrm{K}$ for $V_{g}=2\;\mathrm{V}$).
Actually, the thermal noise of a transistor, $S_{therm}\simeq 4|G+\beta G_m|k_BT_e$ deviates from the standard Johnson-Nyquist formula for a two terminal resistor by an additive correction proportional to the transconductance $G_m$. This effect has been demonstrated in carbon nanotube transistors in Ref.[\onlinecite{Chaste2010apl}]. The abrupt sign change of $G_m$ observed at $V/V_{g}\simeq1.4$ in Fig.\ref{GRS1_overview}-right, actually secures a full compensation of the residual conductance which explains the full suppression of thermal noise observed at $I/I_{sat}\simeq1.1$ in Fig. \ref{bruit}-left.

Assigning the $S_I(I_{ds})$ minima to shot noise we deduce a pinch-off shot noise
$S_{shot}(I_{sat})\simeq 2eI_0\times I_{sat}/\sqrt{I_1^2+I_{sat}^2}$ (dashed blue line),
with $I_0=0.26\;\mathrm{mA}$ and  $I_1=6\;\mathrm{mA}$. It translates into a tiny Fano factor,
$\mathcal{F}=I_{shot}/I_{sat}\simeq I_0/\sqrt{I_{sat}^2+I_1^2}\lesssim 0.04$,
that vanishes for $I_{sat}\gg I_1$.  We regard the presence of shot noise as a proof of the existence of a pinch-off junction,
and attribute its tiny Fano factor to a consequence of strong collimation effect.

\renewcommand{\thetable}{SI-\arabic{table}}

\begin{table}[t]
\centering
\hspace*{-0.5cm}
    \begin{tabular}{| p{1.4cm} |  c |  c | c | c | c | c |  c |  c |  c |  c |  c |c|}
    \hline
    \textbf{Sample} & \textbf{L} & \textbf{W}& \boldmath{$t_{hBN}$} & \boldmath{$t_{Graph}$} & \boldmath{$\mu$} & \boldmath{$R_{c}$} & \boldmath{$A$} & \boldmath{$\sigma_{Z}$} & \large{\boldmath{$\;\;\frac{V_{Z}}{V_{g}}\;\;$}}& \large{\boldmath{$\;\frac{eV_{sat}}{\mu_{s}}\;$}} & \large{\boldmath{$\frac{V_{S}}{V_{g}}$}}&\large{\boldmath{$G_0$}}\\
       &$\;\mu \mathrm{m}\;$ &
    $\;\;\mu \mathrm{m}\;\;$& $\mathrm{nm}$ & $\mathrm{nm}$ & $\;\frac{\mathrm{m}^{2}}{\mathrm{Vs}}\;
    $ & $\;\;\;\Omega\;\;\;$ & -&  $\;\; \mathrm{mS}\;\;$ & -&-  & -&$\mathrm{mS}$\\ \hline\hline
    \textbf{GrS1} & 13 & 17 & 25 & 15 & 6.3 & 80 & 50 & 2.5& 1.25 & 2.6 & $ \;\;0.4 +0.37\; n_{s}\;$ &$0.17\pm0.02$\\ \hline
    \textbf{GrS2} & 10.5 & 15 & 35 & 7 & 13 & 80 & 40 & 1 & 1.7 &3.3 & $ \;\;0.7 +0.30\; n_{s}\; $&$0.18\pm0.01$\\ \hline
    \textbf{GrS3} & 15 & 10 & 42 & 7 & 12 & 120 & 70 & 1 & 1.9  & 2.8 & $ \;\;0.7 +0.53\; n_{s}\; $& $0.18\pm0.02$\\ \hline
    \textbf{AuS1} & 10 & 13.4 & 34 & / & 8 & 80 & 7 & 1.5 & 1.7  & 3.7 & $ \;\;0.4 +0.17\; n_{s}\;$&$0.18\pm0.02$ \\ \hline
    \textbf{AuS2} & 16 & 10.6 & 32 & / & 13 & 80 & 11 & 1 & 1.5  & 4.9 & $ \;\;0.4 +0.18\; n_{s}\;$&$0.18\pm0.01$ \\ \hline
    \textbf{AuS3} & 11.1 & 11.4 & 90 & / & 11 & 95 & 100 & 0.7 & 1.6  & / & $\;\; 0.7 + 1.05\;n_{s}\;$&$0.17\pm0.04$\\ \hline
    \end{tabular}
\caption{Summary of the properties of the device series studied in the main text (GrS3) and in the supplementary (GrS1 and GrS2, AuS1, AuS2, AuS3). Base parameters presented in the first columns include the geometrical properties $L,W,t_{hBN}$, $t_{Graph}$ for the devices on graphite gate, mobility and contact resistance. Also displayed is the maximum value of the voltage gain $A=\frac{\partial V}{\partial V_{g}}$. The next 4 columns correspond to the parameters that are responsible for the unveiling and visibility of the Klein-Schwinger effect in the different devices : $V_{Z}$ is the Zener onset voltage, where a Zener conductance $\sigma_{Z}$ activates. It has to be (notably) higher than the Schwinger voltage $V_{S}$ that triggers the Schwinger conductance, and whose doping dependence is shown in the next-to-last column. The doping-dependent saturation voltage $V_{sat}$, extracted from the dependence $G= G(0) Exp\left[-\frac{V}{V_{sat}}\right]$ in the Klein regime, is a metrics of the Klein collimation and defines the beginning of the conductance gap, later terminated by Zener regime. The last column $G_0$ corresponds to the zero-bias affine extrapolate of the Schwinger pair conductance, as explained in the main text.}
\label{tab:label}
\end{table}

\section{Klein-Schwinger effect in the full device series}
\label{section:dispos}

Besides the sample GrS3 extensively described in the main text, five other samples have been measured and characterized. All of them show consistent Klein-Schwinger effect, albeit with a somewhat lower visibility due to non-optimal geometrical properties or electronic mobility. Yet, the subtraction of the fitted Klein conductance, following the same procedure as in the main text, allows restoring a larger visibility of the 1d-Schwinger conductance for all the devices. A summary of the properties of the five devices can be found in Table \ref{tab:label}. For completeness we have reproduced the data of sample GrS3 analyzed in main text.

\begin{figure}[h!]
\hspace*{-0.5cm}
\centering{}\includegraphics[scale=0.24]{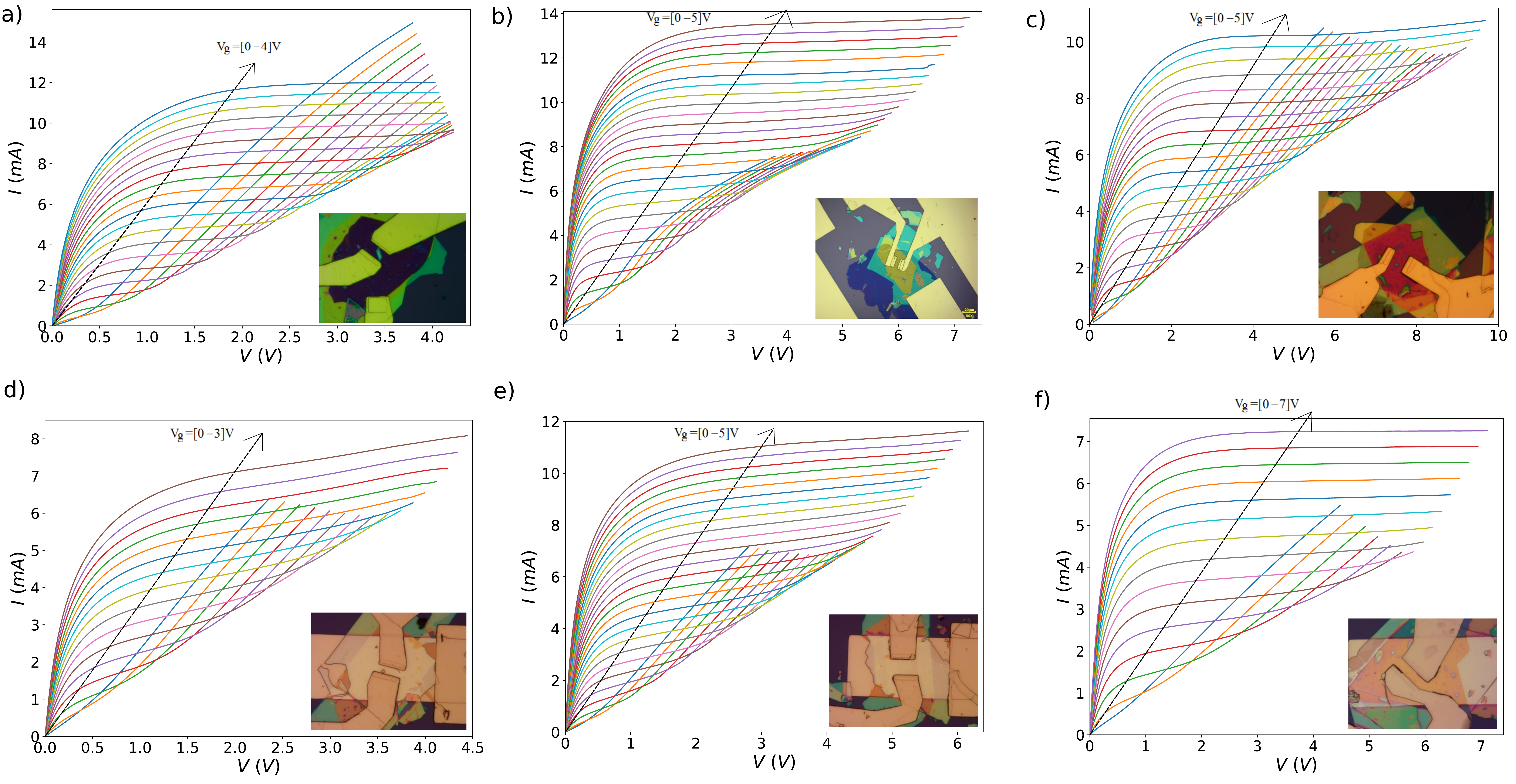} \caption{Current-voltage curves of the device series GrS1 (panel a), GrS2 (panel b), GrS3 (panel c), AuS1 (panel d), AuS2 (panel e) and AuS3 (panel f). Inset of each panel shows an optical image of the device. Actual device dimensions and properties for the series are collected in Table \ref{tab:label}. The applied gate voltages correspond to a charge carrier density at the source ranging from $0$ to $1.5-2.5~ 10^{12}\, \mathrm{cm^{-2}}$ depending on the device.} 
\label{IV6plots}
\end{figure}

The current-voltage curves of the six devices series are shown in Figure \ref{IV6plots}. We always subtract in the bias voltage the voltage drop $R_CI$ across the contact resistance. All the devices exhibit broad current saturation plateaus, with, however, an incomplete saturation (differential resistance peak $G^{-1}\lesssim 5\mathrm{k}\Omega$) for devices AuS1 (panel d) and Au-S2 (panel e). The current saturation plateaus are terminated by the onset of Zener conductance at a Zener field $V_{Z} = 1.25-2 V_{g}$ depending on the device. Detailed values can be found in Table \ref{tab:label}.

\begin{figure}[h!]
\hspace*{-0.5cm}
\centering{}\includegraphics[scale=0.24]{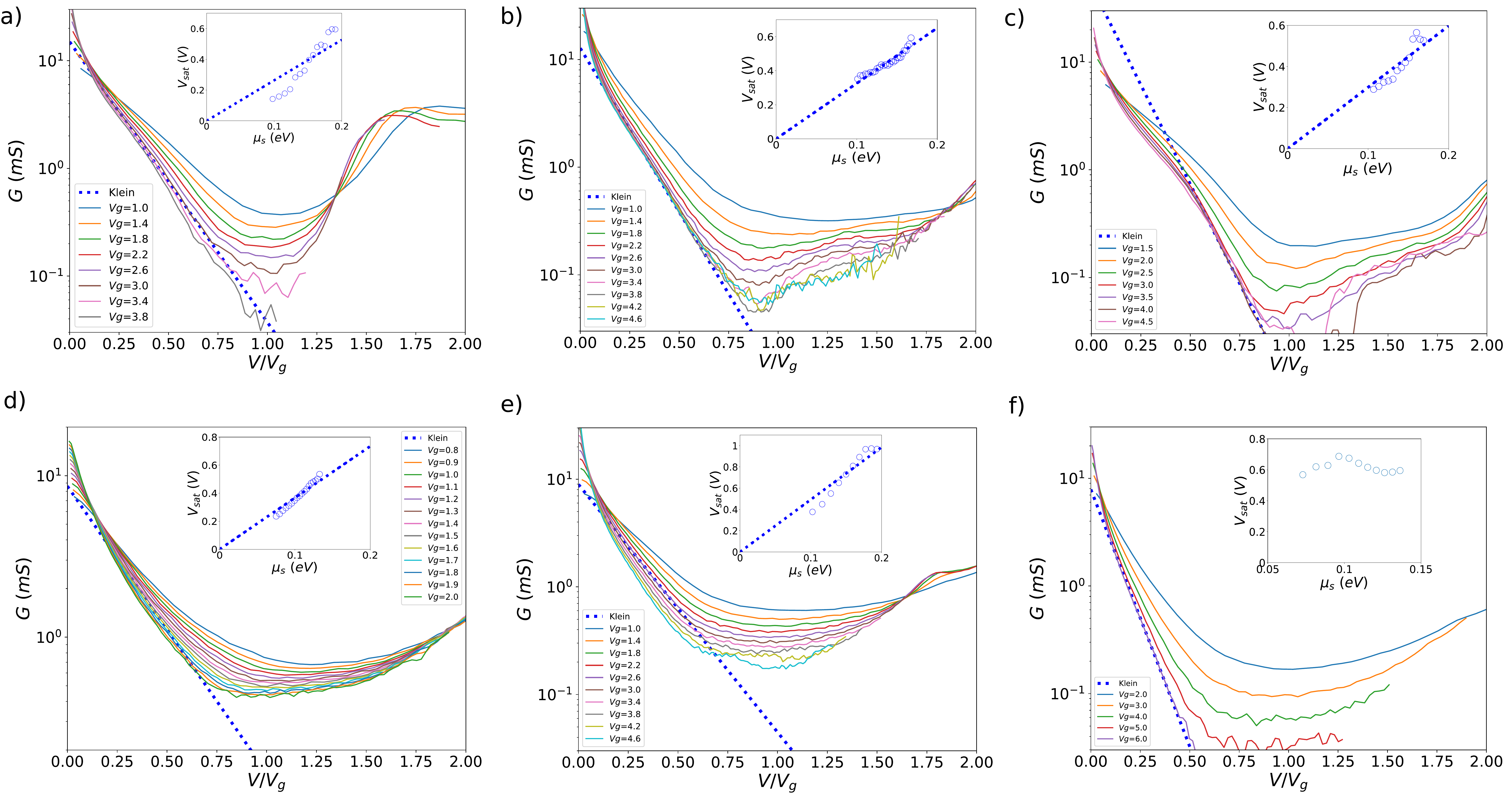} \caption{Semi-log representation of the differential conductance scaling $G(V/V_{g})$, showing the exponential decay of the saturation dependence $G=G(0) exp(-\frac{V}{V_{sat}})$ (dotted blue line) for the device series GrS1 (panel a), GrS2 (panel b), GrS3 (panel c), AuS1 (panel d), AuS2 (panel e) and AuS3 (panel f). Inset of each panel shows the evolution of the saturation voltage $V_{sat}$ as function of the doping, with a linear relation for all devices except AuS3. Numerical values can be found in Table \ref{tab:label}. }
\label{Klein6plots}
\end{figure}

All six devices exhibit consistent Klein collimation, as is described by the semi-log representation of the differential conductance scaling $G(V/V_{g})$ in Figure \ref{Klein6plots}. It shows an exponential decay of the Klein conductance for all samples, with a sample-dependent slope that tunes the depth of the conductance gap. For instance, sample AuS3 (panel f) exhibits a steep conductance decay by three orders of magnitude at pinch-off, with differential conductance dropping below $10\mu \mathrm{S}$, whereas the conductance gap is very shallow in samples AuS1 (panel d) and AuS2 (panel e). The doping-dependent saturation voltage $V_{sat}$, plotted in inset in the different panels, presents consistently a linear relation with the doping $eV_{sat} = \alpha \mu_{S}$ with $\alpha \sim 3$ in the different samples (except for AuS3 where $V_{sat}=Cte$, see Table \ref{tab:label}).

\begin{figure}[h!]
\hspace*{-0.5cm}
\centering{}\includegraphics[scale=0.24]{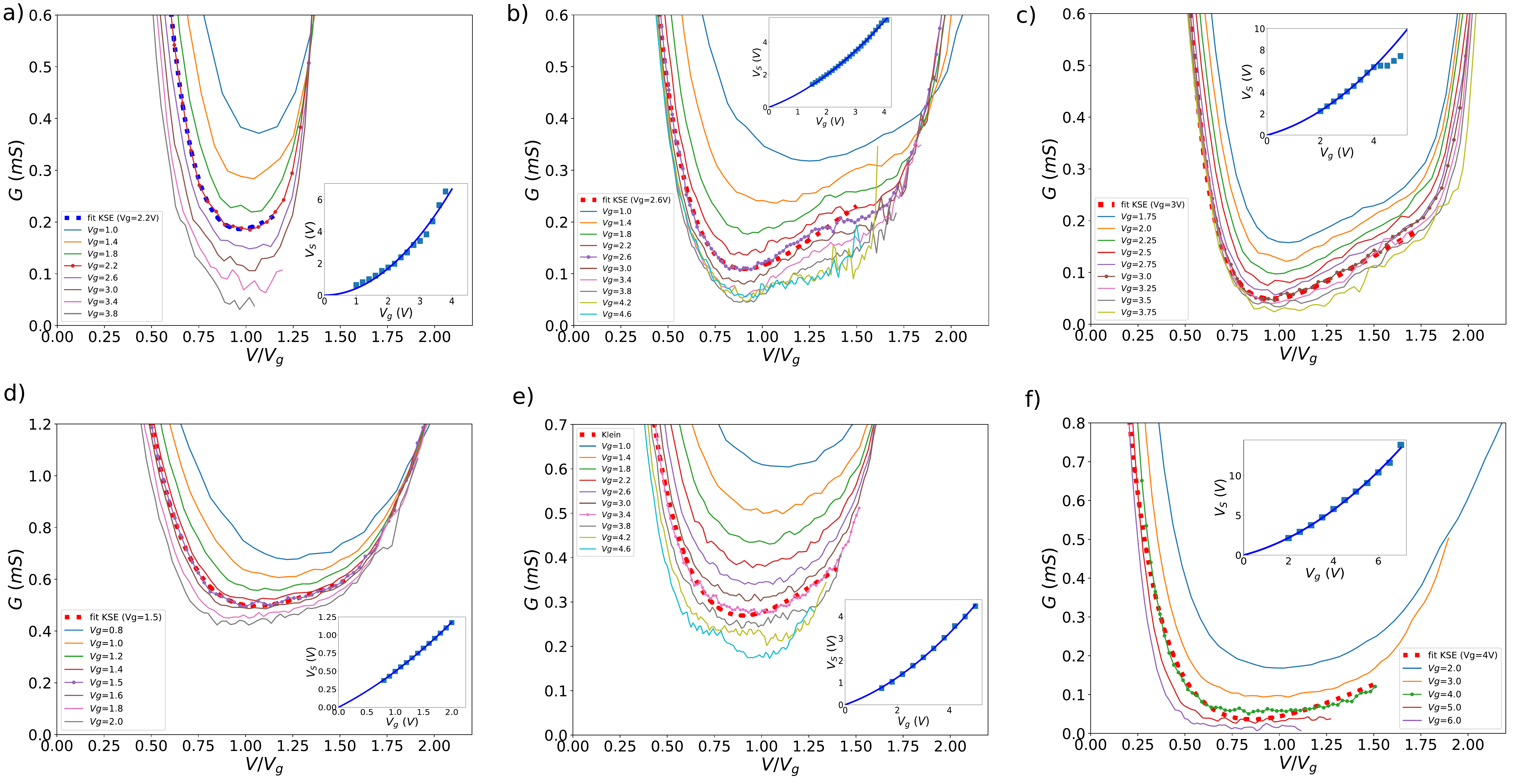} \caption{Klein-Schwinger effect in the sample series GrS1 (panel a), GrS2 (panel b), GrS3 (panel c), AuS1 (panel d), AuS2 (panel e) and AuS3 (panel f). The red dotted lines on each figure show an example of a three-parameter fit $G=G_{S}(V_{S})+G(0)e^{-V/V_{sat}}$ performed in the $V_{sat} \lesssim V \lesssim V_{Z}$ range ; due to different geometrical and electronic properties, the Schwinger visibility is variable on the different devices. Inset in each panel shows the super-linear dependence of the Schwinger voltage as function of gate voltage $\frac{V_{S}}{V_{g}}=a+b n_{s}$ deduced from the fits. The values of the coefficients $a$ and $b$ can be found in Table \ref{tab:label} for each device. }
\label{Schwinger6plots}
\end{figure}

In order to study 1d-Schwinger pair conductance in the full series, Figure \ref{Schwinger6plots} shows a zoom of the differential conductance as function of $V/V_{g}$ in the conductance gap region, where Schwinger conductance appears on top of current saturation. The visibility of Schwinger effect differs on the different devices, yet the 3-parameters KSE fits closely match the data on all of them (dotted lines in the figure). Sample GrS1 (panel a) demonstrates a very narrow conductance gap due to early Zener onset at $V_{Z} \sim 1.25 V_{g}$, hiding Schwinger conductance. Samples GrS2 and GrS3 (panel b and c) are very similar and both exhibit a wide conductance gap due to a complete current saturation and a rejection of the Zener voltage above $1.7 V_{g}$. Due to the Klein conductance $G_{K} \propto W$ and the Zener threshold set by Pauli blocking $V_{Z} \propto L$, sample GrS2 exhibits a narrower conductance gap compared to GrS3. 1d-Schwinger conductance is still clearly visible in the conductance gap.
\\The devices on Au gates have a lower visibility for the Schwinger conductance here. Sample AuS1 (panel d) and AuS2 (panel e), whose properties are very similar (see Table \ref{tab:label}), demonstrate a very shallow conductance gap $G_{min} \sim 0.2-0.4 \textrm{mS}$, making it uneasy to see a direct signature of Schwinger regime via its onset. Sample AuS3 (panel f), with its thicker dielectric $t_{hBN}=90 \textrm{nm}$, exhibits an extremely deep and large conductance gap. However, the large dielectric thickness also shifts the Schwinger voltage towards higher values $V_{S} \gtrsim 1.5 V_{g}$, setting the onset of the Schwinger conductance near the Zener onset and therefore reducing its visibility.
\\The KSE fits performed on the 6 devices give the values of the Schwinger voltage $V_{S}$, which is plotted in the insets as function of the gate voltage $V_{g}$. For analysis of these data, we refer to Figure 3-c in the main text.

\begin{figure}[h!]
\hspace*{-0.5cm}
\centering{}\includegraphics[scale=0.24]{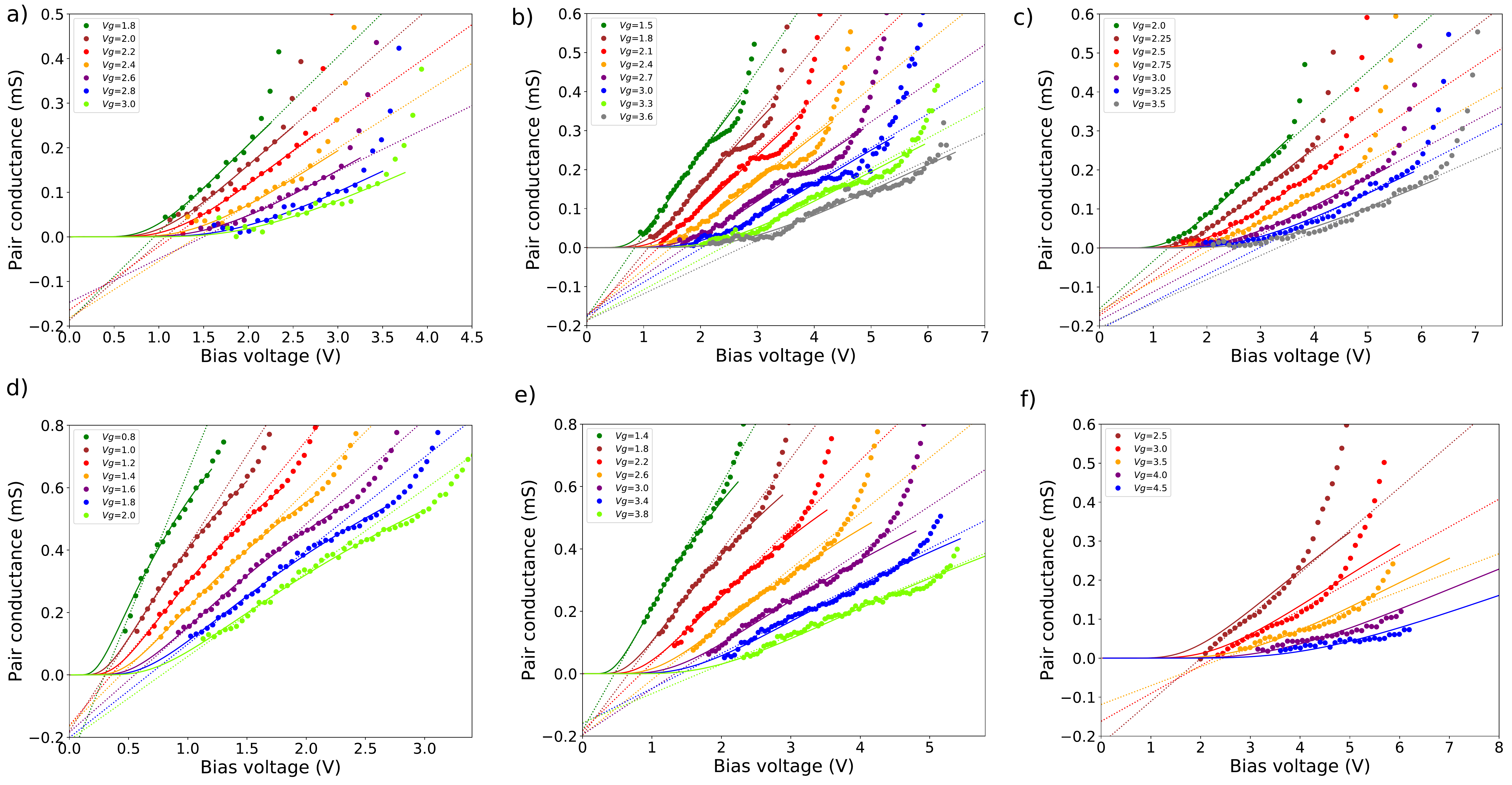} \caption{Pair conductance of the device series GrS1 (panel a), GrS2 (panel b), GrS3 (panel c), AuS1 (panel d), AuS2 (panel e) and AuS3 (panel f), obtained from data by subtracting the fitted Klein conductance. Pair conductance can be fitted using the 1d-Schwinger formula (solid lines) over a device-dependent range of bias, highlighting the presence of Schwinger conductance in all devices. The 'affine approximation' extrapolates at zero bias (dotted lines) collapse onto one universal value $G_{0} \simeq 0.18\ \mathrm{mS}$ which appears doping- and device-independent. Exact values of the extrapolate can be found in Table \ref{tab:label}.}
\label{Pair6plots}
\end{figure}

The visibility of the 1d-Schwinger pair conductance is determined by the combined vanishing of Klein intraband conductance and Zener interband conductance. Using the KSE fits, performed in Klein and Schwinger regimes, the subtraction of the fitted contribution of Klein conductance allows revealing the Schwinger conductance on a broader range of bias, a procedure which has been described in the main text (see Figure 3-b). The resulting pair conductance $G_S(V)$ is plotted for all devices in Figure \ref{Pair6plots}, revealing ubiquitous 1d-Schwinger conductance for the sample series. Agreement with the theoretical formula (solid lines) is excellent, up to a a Zener voltage corresponding to a steep increase of pair conductance. 
\\Whereas the 1d-Schwinger conductance was already clearly visible in samples GrS2 and GrS3 (panels b and c), the subtraction reveals large Schwinger conductance in sample AuS1 and AuS2 (panels d and e) that was previously obscured by Klein conductance exponential decay. Due to the low value of the Schwinger voltage $V_{S} \sim 0.5-0.7 V_{g}$ in these last two devices, Schwinger pair conductance is visible on a wider bias range before Zener conduction sets in, unveiling the non-linearity of Schwinger conductance at high bias voltages which is discussed in the main text (Fig.3-d).
\\The universality of 1d-Schwinger conductance is revealed by the zero-bias 'affine approximation' extrapolate, that gives a unique value $G_{0} \simeq 0.18 \mathrm{mS}$ for all gate voltages and for each of the 6 devices of the series, despite a broad variation of geometrical parameters and electronic mobility. This value closely follows the theoretical expectation $G_{0} = 0.186 \mathrm{mS}$ from Schwinger formula.

Unveiling the visibility of Klein-Schwinger effect is thus an optimization task of the different geometrical parameters; the five devices in this Supplementary Information Section were in fact used to set the optimal visibility properties of the sample GrS3, which is used as a demonstrator in the main text. Prerequisites are a low contact resistance and a very high electronic mobility $\mu \gtrsim 10 \mathrm{m}^{2}\mathrm{V}^{-1}\mathrm{s}^{-1}$. The hBN optimal thickness is a trade-off: thin layers cause dielectric breakdown problems and confine $V_S$ below pinch-off, generating an early Schwinger conductance onset that can only be revealed by subtraction. Thick layers shift the Schwinger voltage upwards, eventually above the Zener onset terminating the conductance gap. Klein-Schwinger effect is more easily visible in long channel transistors, due to the Pauli-blocked Zener voltage, and devices with a smaller width favor the Klein-collimation and the appearance of 1d-Schwinger effect.

\section{Relation between Schwinger voltage, doping and dielectric thickness}

\begin{figure}[h!]
\hspace*{-0.5cm}
\centering{}\includegraphics[scale=0.7]{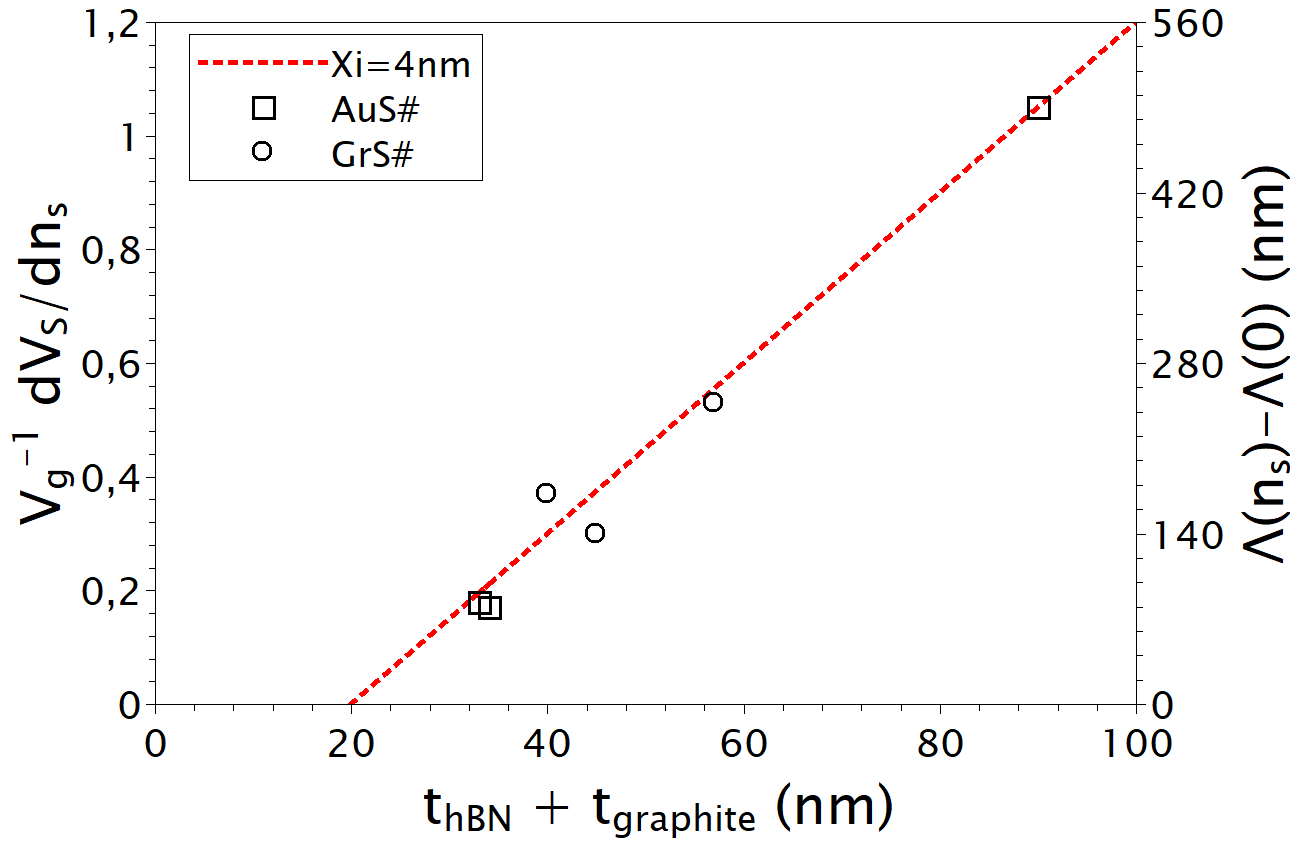} \caption{Slope extracted from the linear increase of the ratio $\frac{V_{S}}{V_{g}}$ as function of doping $n_{s}$ (left axis), plotted with respect to the sum of hBN and graphite thicknesses for the 6 devices. Right axis gives the associated value of the junction length $\Lambda(n_{s})$. The red dotted line is a linear fit corresponding to $\xi = 4\mathrm{nm}$.}
\label{thickness}
\end{figure}

The variety of devices studied in Section \ref{section:dispos} allows us to analyze the dependencies between Schwinger voltage, doping and dielectric thickness.
Figure 3-c of the main text, and the forelast column of Table \ref{tab:label}, show the evolution of the ratio between Schwinger voltage and gate voltage $V_{S}/V_{g}$ as function of doping $n_{s}$ for the 6 devices. They consistently show a linear increase with respect to $n_{s}$ that we can attribute to a Coulomb-repulsion effect from the interactions, with a slope that is device-dependent. This slope is plotted as function of the thickness of the hBN dielectric in Figure \ref{thickness}. For the case of devices on graphite backgates, the hBN thickness has been replaced by the sum of $t_{hBN}$ and the graphite thickness $t_{Graph}$, whose values are given in Table \ref{tab:label}. This relies on the existence of a Debye length for neutral graphite that we approximate to be the total thickness of the thin graphite flake. The slope exhibits an affine increase as function of hBN thickness, and extrapolates to zero for $t_{hBN} \sim 20\mathrm{nm}$. This value is on the order of the Fermi wavelength for typical doping range $n_{s}=1-2 ~10^{12} \mathrm{cm}^{-2}$, which acts as a cut-off for interaction effects. 

Let us make the assumption that the Schwinger gap $\Delta_{S} \sim \mu_{s}$ which is the only energy scale of the problem. Then, the dielectric thickness $t_{hBN}$ being the only lengthscale of the problem, the doping dependence of the junction length $\Lambda (n_{s}, t_{hBN})$  can be cast in the form $\frac{\Lambda}{t_{hBN}}=\frac{V_{S}/E_{S}}{V_{g}/E_{g}}= 4 \alpha_{g} (\frac{\mu_{s}}{\Delta_{S}})^{2} \frac{V_{S}}{V_{g}} \simeq 2.8 \frac{V_{S}}{V_{g}}$ where $\alpha_{g} = \frac{e^{2}}{4\pi \epsilon_{0} \epsilon_{hBN} \hbar v_{F}} = 0.70$ is the graphene fine structure constant, taking the high-field hBN permittivity $\epsilon_{hBN}\simeq 3.1$ [\onlinecite{Pierret2022MatRes}].

The measured evolution of $\frac{V_{S}}{V_{g}}(t_{hBN}, n_{s})$ described in Figure 3-c of the main text and Figure \ref{thickness} can thus be cast into the junction length power expansion $\Lambda \simeq a t_{hBN}+ \xi n_{s} t_{hBN}^{2} $. Numerical values of $\Lambda$ are indicated on the right axis in Figure \ref{thickness} and yield $\xi \simeq 4\mathrm{nm}$, which corresponds to the typical interaction radius per electron , quantifying the doping-induced dilation of the junction length.

The direct consequence of these dependencies is that the 1d-Schwinger conductance becomes hardly visible for devices with thick dielectric such as AuS3, where the Schwinger voltage exceeds the Zener onset voltage even at moderate doping. On the contrary, devices with thin dielectric show an early Schwinger conductance over a large range of doping.

\begin{figure}[h!]
\hspace*{-0.5cm}
\centering{}\includegraphics[scale=0.5]{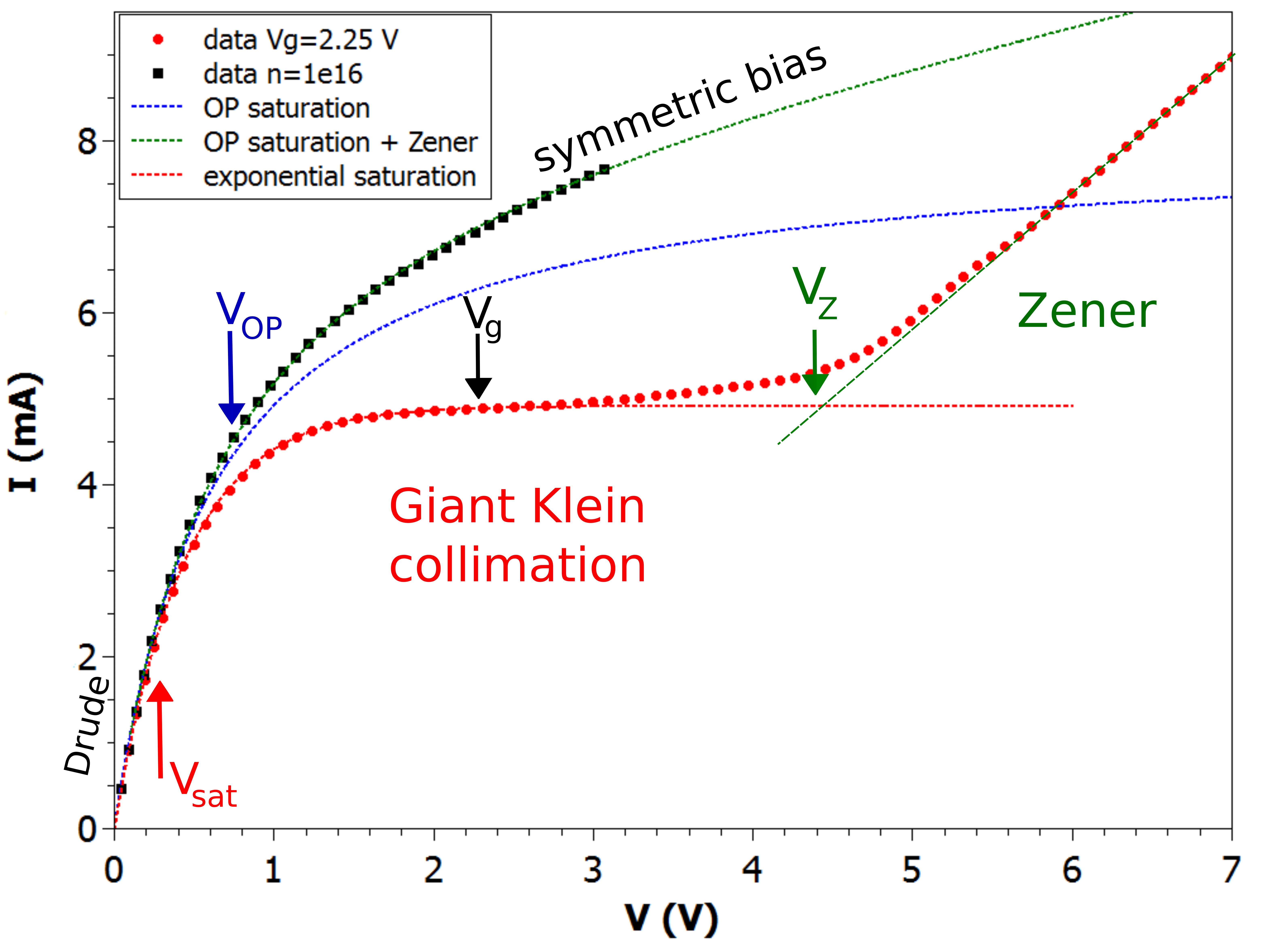} \caption{Comparison between pinchoff and pinchoff-free regimes in sample GrS3 at the carrier density $n_{s}=10^{12}\;\mathrm{cm}^{-2}$. 
Black dots correspond to measurements at fixed $V_{g}=2.25\mathrm{V}$, whereas red squares correspond to a measurement along constant density line $V_{g}(V)= V_{g} +0.4V$ with $V$ the bias voltage, for the same density. The current saturation in the pinch-off free case can be fitted using saturation by optical phonons ( blue dotted line, with the green dotted line adding the contribution of interband Zener conductance) with $\epsilon_{sat} = 91~\mathrm{meV}$, showing that the bottleneck of current saturation at constant $V_{g}$ is not due to optical phonons but to the pinch-off. Red dotted line is a fit for an exponential current saturation at pinch-off.  }
\label{compar}
\end{figure}

\section{Pinch-off of high-mobility graphene transistors}

We first start with a comparison between pinchoff and pinchoff-free characteristics.  As a general rule, finite biases induce both doping and voltage drops in transistors, which are proportional to the source-drain voltage. The difference between pinchoff and pinchoff-free cases is about their spatial distributions: in the pinchoff case, they are confined at the drain side, whereas in the pinchoff-free case they are distributed along the channel. In the pinchoff-free diffusive regime of bottom/top gated transistors, the bias voltage entails a uniform doping gradient along the channel from $n_s=C_g(V_g)/e$ at the source to $n_d=C_g(V_g-V)/e$ at the drain. The pinchoff-free regime is achieved by applying a bias-dependent gate voltage $V_{g}(V)= V_{g}(0) +aV$ with $a\sim0.5$ mimicking a symmetric $\pm V/2$ bias. Experimentally $a\simeq0.4$ is adjusted to obtain a bias-independent charge neutrality point position. In the pinchoff case, most of the doping gradient is localized at the drain in a pinchoff junction, leaving the doping quasi uniform in the channel.   

Fig.\ref{compar} shows the   contrasted pinchoff (red circles) and pinchoff-free (black circles) characteristics of sample GrS3, for a carrier density $n_{s}=10^{12}\;\mathrm{cm}^{-2}$. Both deviate at $V_{sat}$, with the pinchoff I-V (asymmetric bias)  showing an exponential current saturation (red dotted line), as opposed to the pinchoff-free (symmetric bias) I-V which shows strong non-linearities but does not saturate. These non-linearity is qualitatively understood (green dotted line) in terms of optical-phonon-induced
velocity saturation (blue dotted line) with a characteristic voltage $V_{OP}$ , and interband Zener tunneling as captured by the formula
\begin{equation}
I=n_se\left[\frac{\mu}{1+V/V_{OP}}+\sigma_Z\right]\frac{W}{L}V\quad ,
\label{symmetric}
\end{equation}
with $\mu\simeq12\;\mathrm{m^2/Vs}$, $V_{OP}\simeq0.65\;\mathrm{V}> V_{sat}$, and $\sigma_Z\simeq0.5\;\mathrm{mS}$.
In the following we go beyond that phenomenological description by solving a simple 1d-model of the electrostatic potential distribution in the channel.

\begin{figure}[h!]
\hspace*{0cm}
\centering{}\includegraphics[scale=0.4]{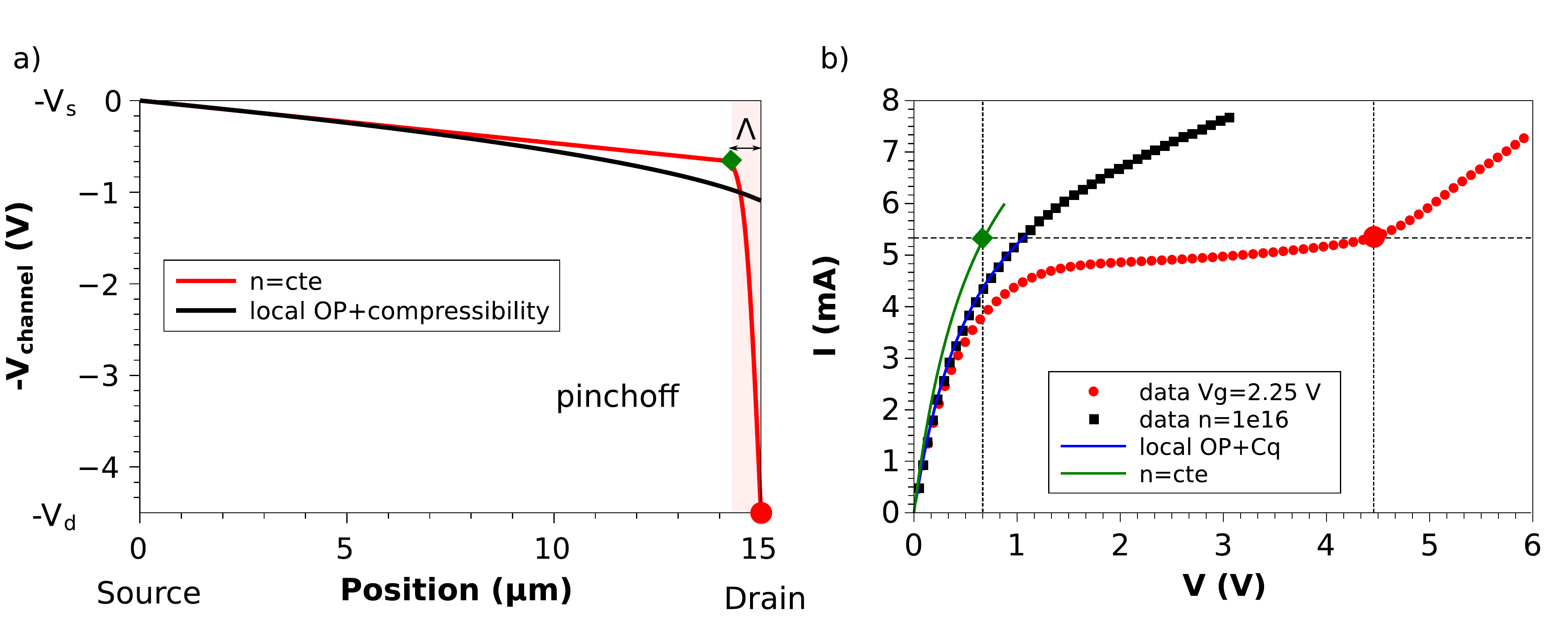} \caption{Panel a) Electrostatic profile in the constant density case and in the pinchoff-free case, with an excess potential drop at the drain. Both profiles are computed for a 5.4mA current. Panel b) Current-voltage curves in the two biasing regimes. The green black line corresponds to the constant density model of the channel described in the text, and the blue-line fit is performed with the microscopic model presented in the text. The horizontal dashed line corresponds to the 5.4mA current of panel a. The geometrical construction allowing to recover the potential drop in the channel and at pinchoff is indicated on the figure by the bold symbols. }
\label{SI10}
\end{figure}

Focusing on the pinchoff-free case, we estimate the electrostatic potential profile by solving a local microscopic model of transport including: velocity saturation by OPs, electronic compressibility, and electrostatic equilibrium in the presence of a local gate. Dealing with current lass than or on the order of pinchoff saturation currents, we neglect the Zener tunneling contribution (see Fig.\ref{compar}).

The current density writes
$$
J=\mu(E) n\partial_{x}\mu_{c}^{*},
$$
where $\mu(E)=\mu(0)/(1+|E|/E_{OP})$ is the electric-field-dependent mobility accounting for OP velocity saturation  in Eq.(\ref{symmetric}), $\mu(0)$ is the zero-bias  mobility, $\mu_{c}^{*}=\mu_c(x)-e V_c(x)$ and  $\mu_c(x)$  the electrochemical and chemical potentials. $V_c$ is the electrostatic potential distribution of interest in this section.  For monolayer graphene, $\mu_{c}^{*}-\mu_{g}^{*}=\hbar v_{F}\sqrt{\pi n}+\frac{e^{2}}{c_{g}}n$ where $v_F$ is the Fermi velocity,  $c_{geo}$ is the areal gate-channel geometric capacitance, $n(x)$ is the carrier density, and $\mu_{g}^{*}=-eV_g$ is the uniform electrochemical potential of the gate. Additionally, electrostatics imposes $V_{c}-V_{g}=\frac{-e}{c_{geo}}n(x)$, so that the differential equation for the channel voltage writes :
\begin{equation}
J=\frac{I}{W}=-\frac{\mu_{0}}{1+|\partial_{x}V_{c}|/E_{OP}}c_{geo}(V_{c}-V_{g})\ \left\{ \frac{\hbar v_{F}}{e}\sqrt{\frac{\pi c_{geo}}{e}}\frac{1}{2\sqrt{V_{g}-V_{c}}}+1\right\}\partial_{x}V_{c}\quad ,
\label{Manu}
\end{equation}
where $E_{OP}=V_{OP}/L$ is the OP saturation electric field. 

We  numerically solve this equation for currents in the range $[0-5]$mA imposing $V_d+V_s=0,$ and $V_g=-2.25\ V$ to match the experimental conditions and find perfect agreement (solid blue line in Figure \ref{SI10}b) with the IV experimental response for  $\mu(0)=12.3\ \mathrm{m^2.V^{-1}.s^{-1}}$, $E_{OP}=43\ \mathrm{mV/\mu m}$. 
Using this set of parameters, we next compute the channel potential profile at the current saturation in the pinchoff-free case for a $5\mathrm{mA}$ current, which is plotted  as a black line in Figure \ref{SI10}a. The total channel voltage drop, $V_c(L)=1\;\mathrm{V}$ (see large black dot in Fig.SI-\ref{SI10}b). 

To model the pinchoff case, we assume a constant carrier density  in the channel, which implies a constant chemical potential and uniform in-plane electric field. 
As a consequence, the differential equation (\ref{Manu}) simplifies as :
$$J=-\frac{\mu_{0}}{1+|\partial_{x}V_{c}|/E_{OP}}n\ 
\partial_{x}V_{c}$$.

The electrostatic energy profile in this case for a $5.4\mathrm{mA}$ current is plotted in Fig. \ref{SI10}a (red line) up to the drain-collimation junction where potential is assumed to drop abruptly down to the applied voltage $V=4.5\;\mathrm{V}$. 
The total channel voltage drop, $V_c(L-\Lambda)\simeq0.5\;\mathrm{V}$ remains small with respect to the total bias $4.5\;\mathrm{V}$, ass assumed in the main text.

Fig. \ref{SI10}b summarizes the conclusions of our potential distribution analysis. At a given bias current, the respective channel and junction potential drops can be deduced by the geometrical construction displayed in the figure.